\documentclass[ twocolumn,final
amsmath,amssymb,
pre,
]{revtex4-2}

\usepackage{graphicx}
\usepackage{dcolumn}
\usepackage{bm}

\usepackage{amsmath}
\usepackage{epstopdf}
\usepackage[T1]{fontenc}
\usepackage[utf8]{inputenc}
\newcommand{\includegraphic}[5][,]{%
	\setbox1=\hbox{\includegraphics[#1]{#2}}
	\leavevmode\rlap{\usebox1}
	\rlap{\hspace*{#4}\raisebox{\dimexpr\ht1-#5\baselineskip}{\normalsize{#3}}}
	\phantom{\usebox1}}

\begin{document}
	
	\preprint{APS/123-QED}
	
	\title{Smallest chimeras under repulsive interactions}
	
	\author{Suman Saha}
	\email{ecesuman06@gmail.com}
	\affiliation{National Brain Research Centre, Manesar, Gurugram 122051, India}
	\author{Syamal Kumar Dana}%
	\email{syamaldana@gmail.com}
	\affiliation{National Institute of Technology, Durgapur 713209, India
	}%
	\affiliation{Division of Dynamics, Lodz University of Technology, Lodz 94-924, Poland 
	}%
%
%
	
	\date{\today}
	
	\begin{abstract}
We present an exemplary system of three identical oscillators in a ring interacting  repulsively to show up chimera patterns. 		
The dynamics of individual oscillators is governed by the superconducting Josephson junction. Surprisingly, the repulsive interactions establish a symmetry of compelete synchrony in the ring, which is broken with increasing  interactions when the junctions pass through  serials of asynchronous states (periodic and chaotic), but finally emerge into chimera states. The chimera pattern appears in chaotic rotational motion of the three junctions when  two junctions evolve coherently while the third junction is incoherent. 
For larger repulsive coupling, the junctions evolves into another chimera pattern in a periodic state when two junctions remain coherent in rotational motion and one transits to incoherent librational motion. This chimera pattern is sensitive to initial conditions, in the sense, that the chimera state flips to another pattern when two junctions switch to coherent librational motion and the third junction remains in rotational motion, but incoherent. The chimera patterns are detected by using  partial and global error functions of the junctions while the librational and rotational motions are identified by a libration index. All the collective states, complete synchrony, desynchronization and two  chimera patterns, are delineated in a parameter plane of the ring of junctions, where the boundaries of complete synchrony are demarcated by using the master stability function. 
	\end{abstract}
	
	\maketitle
	
	
\section{Introduction}

Chimera states \cite{parastesh2020chimeras, majhi2019chimera, wang2020brief, abrams2004chimera, sethia2008clustered, martens2013chimera, laing2009chimera, gopal2014observation, hagerstrom2012experimental, hart2019delayed, omelchenko2013nonlocal} became a paradigm of collective phenomena in dynamical systems that started with the first report by Kuramoto et al \cite{kuramoto2002coexistence} of two coexisting synchronous and asynchronous groups of identical phase oscillators arranged in a ring and coupled in a non-local fashion.  The main question was how the symmetry in a completely synchronous ensemble of identical oscillators breaks into two clusters, one synchronous and another asynchronous groups. How  stable  is this chimera state? In the begninning, it was apprehended that the chimera state is a transient behavior; the transient time increases with the size of a network \cite{wolfrum2011chimera, rosin2014transient}. Later it has been  established that chimera states are possible stable states \cite{omel2018mathematics, laing2019dynamics, pecora2014cluster} in an enemble of identical oscillators and with a symmetry in the connectivity matrix or the topology of a network. By this time, this phenomenon has been widely explored in single-layer \cite{ omelchenko2013nonlocal, parastesh2020chimeras, majhi2019chimera, wang2020brief,
	abrams2004chimera, sethia2008clustered, martens2013chimera, laing2009chimera, gopal2014observation, hagerstrom2012experimental, hart2019delayed}, multilayer networks \cite{maksimenko2016excitation, ghosh2016emergence, sawicki2018synchronization, ruzzene2020remote} and 3D networks \cite{maistrenko2015chimera, kundu2019chimera, kasimatis2018three} with different forms of chimeras such as traveling chimera \cite{omel2019traveling, dudkowski2019traveling, alvarez2021traveling}, spiral chimera \cite{martens2010solvable, gu2013spiral}. Various dynamical models \cite{bera2016chimera, banerjee2016chimera, saha2019chimera, hizanidis2015chimera} with different coupling schemes \cite{bera2017chimera, meena2016chimera}, and global coupling \cite{sethia2014chimera, yeldesbay2014chimeralike,hens2015chimera, mishra2015chimeralike}, have been used for observing  chimera patterns. The concepts of amplitude chimera \cite{hens2015chimera, zakharova2016amplitude, banerjee2018networks} and amplitude death chimera \cite{zakharova2014chimera, banerjee2015mean}  have been introduced. All the examples of chimera states revolve around the symmetry breaking of a complete coherent state into coherent and incoherent groups. This perception has been extended further to the observation of two coexisting subgroups of in-phase and antiphase  oscillators \cite{maistrenko2017smallest}, which was also referred to as a chimera state. The necessary requirement of a large set of oscillators for chimera states to observe has also been  relaxed with a smaller size of the network: Chimera patterns emerge in a set of  4-oscillators \cite{meena2016chimera, hart2016experimental, senthilkumar2019local} and even 3-oscillators \cite{maistrenko2017smallest, wojewoda2016smallest}. In the laser system \cite{hart2016experimental}, delay in coupling has been used  for the observation of chimera states in 4-oscillators. A modified Kuramoto phase oscillators with inertia \cite{maistrenko2017smallest} were used to demonstrate both in-phase and anti-phase chimeras in 3-oscillators, and later confirmed in experiments with three attractively coupled pendula \cite{wojewoda2016smallest}.  In almost all the reported studies, attractive coupling has been used for the observation of chimeras while a few examples are found to use a combination of both attractive and repulsive coupling \cite{hens2015chimera, mishra2015chimeralike} to originate chimera states in globally coupled oscillators. 
\par We focus here on the role of repulsive interactions in the origin of chimera pattern in a small ensemble of dynamical units. As reported earlier \cite{mishra2017coherent, ray2020extreme}, chimera states may emerge in a large ensemble of globally coupled Josephson junctions under repulsive interactions. Indeed, the repulsive coupling can induce chimera states in a smallest set of three Josephson junctions arranged in a ring as we demonstrate here. Interestingly, a single Josephson junction shows typical neuron-like spiking and bursting behaviors \cite{dana2006spiking, mishra2021neuron, hongray2015bursting, hens2015bursting} which is one of the reasons that encourage us to investigate the role of repulsive interactions (inhibitory in the sense of neuronal interaction) in a ring of 3-junctions. A single Josephson junction is represented by a resistance-capacitance-shunted-junction (RCSJ) circuit \cite{dana2001chaotic, dana2006spiking, mishra2021neuron}, which remains in an excitable state until an external bias current is applied across the junction. The junction shows spiking limit cycle oscillation for bias current above a critical value. The dynamics of the junction also depends upon a damping-like parameter that is related to the shunted resistance and capacitance of the junction. It shows a bistability region, in parameter space, where limit cycle coexits with a steady state for a range of low damping. The RCSJ model can be represented by a second order phase dynamics of the junction that governs the current flow through the jucntion and the voltage drop across it. The dynamics of the jucntion is rotational when the trajectory of the junction makes a complete rotation around a cyclindrical surface like an inverted pendulum thereby orginating a large amplitude spiking oscillation. On the other hand, the junction may show, for appropriate choice of parameters, a librational motion  when the trajectory of the junction dynamics is restricted to a small region of the cylindrial phase space. This  appears like the small amplitude oscillation in a pendulum.

\par 
A ring of three identical oscillators represents an all-to-all globally coupled network and hence perfectly symmetric. The junctions in the ring show an $2\pi/3$ out-of-phase motion in cyclic order in time, but emerge into a state of complete synchrony (CS) for  repulsive interactions above a critical value. When the repulsive interactions in the ring is increased further, the symmetry is broken with the emergence of  a sequence of desynchronous states followed by chimera patterns, (1) two junctions are in complete  synchrony (CS) in a state of rotational motion and one in incoherent rotational motion when they are all in chaotic state, (2) two junctions in CS in rotational motion and one junction in incoherent libration when the dynamics of the junctions is periodic; this chimera pattern flips to two coherent junctions in libration and one in incoherent rotation for a change in initial conditions. We numerically delineate the different collective states  in a two parameter plane of the junctions using coherence measures and libration index, and use the master stability function (MSF) to demarcate the stable region of CS in parameter plane.
\begin{figure}
	\centering
	\includegraphic[width=2in,height=1.2in]{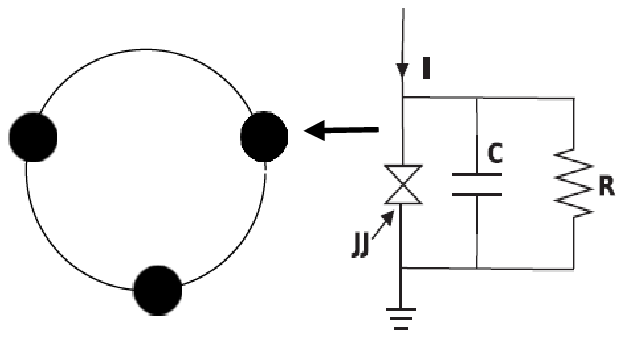}{(a)}{0pt}{0}
	\includegraphic[width=2in,height=1.2in]{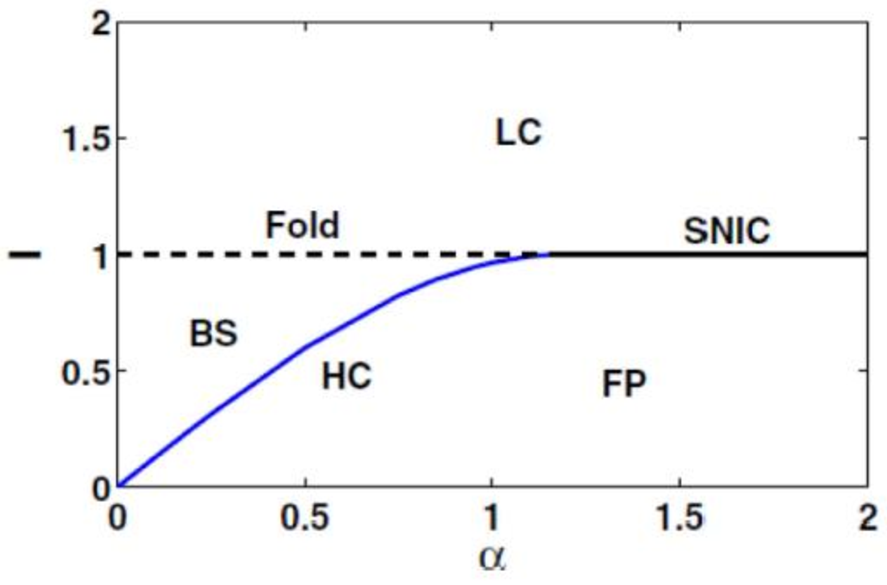}{(b)}{0pt}{0}
	\caption{(a) A ring of three Josephson junctions (JJ). A schematic diagram of one RCSJ ciruit shown at right that represents the dynamics of each node in black circle. (b) Phase diagram of one RCSJ dynamics in a $\alpha-I$ parameter plane. Normalized critical current $I_c$=1.0. The junction shows limit cycle (LC) oscillation for $I>I_c=1.0$. The junction has a bistability (BS) region bounded by a fold bifurcation line (horizontal dashed line) and a homoclinic (HC) bifurcation curve (blue line). A stable fixed point (FP) exists for $I<1.0$ which transits to LC state via saddle-node-invariant circle (SNIC) bifurcation (solid horizntal line).   }
	\label{fig1}
\end{figure} 
\section{Three Josephson junctions in a ring} 
A schematic diagram of the ring of 3-junctions is shown in Fig. \ref{fig1}(a). The RCSJ circuit is shown at right that represents the dynamics of each node. The dynamics of the ring of junctions is represented by the phase dynamics  $\theta_i$  and voltage  $v_i$ $(i=1, 2, 3)$ across the $i^{th}$ junction as given by \cite{mishra2017coherent, josephson1962possible},
\begin{eqnarray}
	\dot{\theta_i} &=& v_i,  \\  
	\dot{v_i} &=& I-sin\theta_i- \alpha v_i +\varepsilon(v_{i-1}-2v_i +v_{i+1}),  \nonumber
	\nonumber
\end{eqnarray}
where $\alpha=[h/2\pi e I^2RC]^{1/2}$ is considered as the damping parameter, $h$ is planck's constant and $e$ is the electronic charge, and $R$ and $C$ are the resistance and capacitance of a junction, respectively.  $I$ is the external bias current normalized by the critical current $I_c$ at each junction. The interaction between the nodes is established through junction voltages with a coupling strength $\varepsilon$, which is considered repulsive (negative) to observe our targeted chimera patterns. Figure \ref{fig1}(b) shows a phase diagram in a $\alpha-I$ plane for a single junction. The single junction remains excitable (stable fixed point, FP) for bias current $I<I_c$ and 
transits to limit cycle oscillation via saddle-node-invarinat circle (SNIC) bifurcation  for $I>I_c$ and $\alpha>1.19$. In a lower range of $\alpha<1.19$ and $I<I_c$ the junction shows bistability (BS) when limit cycle coexists with a steady state. The bistable state transits to limit cycle oscillation via fold bifurcation when $I>1.0$. The bistability region is bounded by fold bifurcation (dashed horizontal line) and homoclinic (HC) bifurcation  (blue curve) in the $\alpha-I$ parameter plane.  For our observation of chimeras, we make a choice of parameters $I>I_c$ and $\alpha>1.19$ so as to obtain limit cycle oscillation (rotational motion) in all three junctions in uncoupled state and keep them away from the bistable region to avoid further complexity in dynamics.
\section{Coherence and libration index }	
To test complete synchrony in three coupled jucntions, we first calculate the error functions between all the pairs of nodes,
\begin{align}
	&e_{12}=\sqrt{\left<(v_1-v_2)^2\right>};\;\;
	e_{13}=\sqrt{\left<(v_1-v_3)^3\right>};\nonumber\\
	&e_{23}=\sqrt{\left<(v_2-v_3)^2\right>},
\end{align}
where, $\left<.\right>$ indicates time average. We check all the three error functions, simultaneously and detect a chimera state when two of the partial error functions are zero (say, $e_{12}$=0 and $e_{13}$=0).

The global error variable is calculated,
\begin{align}
	err=(e_{12}+e_{13}+e_{23})/3
\end{align}
For complete synchrony in the network $err=0$, and $err\neq0$ indicates partial synchrony or incoherence. We use a libration index \cite{mishra2017coherent} to identify the rotational and librational motion of the junctions in different collective states, 
\begin{align}
	LI=\frac{1}{3}\sum_{j=1}^{3}\Theta_j
\end{align}
where, $\Theta_j=\Theta(\delta-m_j)$. $\Theta(.)$ is the Heaviside step function, where $\delta$ is an arbitrarily chosen small threshold and $m_j=1-0.5*\left(max(cos[\theta_j(t)])-min(cos[\theta_j(t)])\right)$. The libration index becomes $LI = 0$ for oscillators in libration and $LI = 1$ when they are in rotational motion. A value of $0<LI<1$ indicates the coexistence of librational and rotational motions in the ring of junctions.
\par
\begin{figure}[!h]
	\centering
	\includegraphics[height=6.5cm, width=7cm]{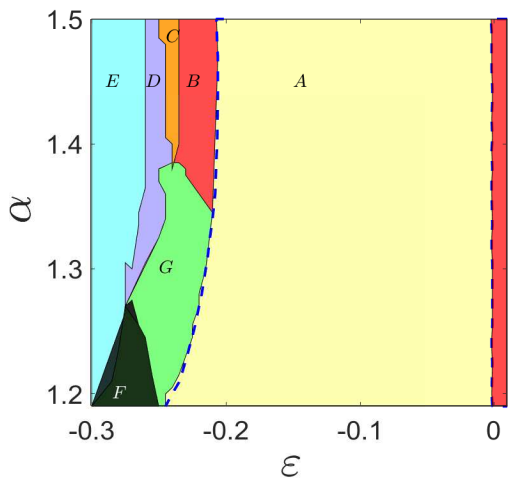}
	\caption{Phase diagram in a $\alpha-\varepsilon$ plane of a ring of three identical Josephson junctions. A large range of CS (A, yellow) is seen for repulsive coupling,  asynchronous  limit cycle dynamics (B, red) and chaos (C, orange) and with a significant range of rotational chimera (D, purple) and chimera of mixed rotational and libration motion (E, cyan) in a broad range of $\alpha>1.2$ for stronger repulsive interactions.  Stability of CS is lost for both attractive coupling ($\varepsilon>0$, red region) and larger repulsive coupling as marked by blue dashedlines at both ends of the CS (A, yellow) region. There are regions of mixed states F (black) and G (green) of coexisiting chimera and desynchronization. Parameter $I=1.5$. }
	\label{fig2}
\end{figure}

 For demarcation of the stable CS state in parameter plane of the junctions in Fig.~\ref{fig2}, we numerically estimate the MSF \cite{pecora1998master} of identical junctions by using the variational equation of the ring of junctions. 
The dynamics of the $i^{th}$ node is $\dot{\bf x}_i={\bf F(x_i)}+\varepsilon\sum_{j}G_{ij}{\bf H(x_j)}$, 
where the state vector ${\bf x}=({\bf x}_1,{\bf x}_2,\cdots,{\bf x}_n)$, and  ${\bf F(x)}=[{\bf F (x_1)},{\bf F(x_2)},\cdots,{\bf F (x_n)}]$, and
${\bf H(x)}=[{\bf H(x_1)},{\bf H(x_2)},\cdots,{\bf H (x_n)}]$ is the coupling matrix and ${\bf G}$ be the adjacency matrix $\{G_{ij}\}$, then
\begin{align}
	\dot{\bf x}={\bf F(x)}+\varepsilon{\bf G}  \otimes {\bf H(x)}, \label{eq5}
\end{align}
where, $\otimes$ denotes the direct product. The variational equation of Eq. (\ref{eq5}) by assuming 
$\xi_i$ as a small perturbation in  the $i^{th}$ node when $\xi=(\xi_1,\xi_2,\cdots,\xi_n)$, ($i=1,2,..n)$ is given by,
\begin{align}
	\dot{\xi}=[{\bf 1_n} \otimes D{\bf F}+\varepsilon {\bf G}\otimes D {\bf H}]\xi
\end{align} when ${\bf H}$ is just a matrix ${\bf E}=(0\;\; 0; 0\;\;1)$, implying coupling in the second variable and $D {\bf H}={\bf E}$. By a block diagonalization of the variational equation, each block will be of the form
\begin{align}
	\dot{\xi_k}=[ D{\bf F}+\varepsilon \gamma_k D {\bf H}]\xi_k, \label{eq7}
\end{align}
where $\gamma_k$ is the eigenvalue of ${\bf G}$, $k=0,1,2,\cdots,n-1$. For $k=0$, we have the variational equation for the synchronization manifold $\gamma_0=0$. The Jacobian matrices $D{\bf F}$ and $D{\bf H}$ are the same for each block, since they are evaluated on the synchronized state.  Thus, for each $k$, the form of each block in Eq. (\ref{eq7}) is same with only the scalar multiplier $\varepsilon\gamma_k$ differing.  This leads us to the following formulation of the master stability equation and the associated MSF: We calculate the maximum Lyapunov exponents $\lambda_{max}$ for the generic variational equation
\begin{align}
	\dot{\zeta}=[D{\bf F} +(\alpha+i\beta)D{\bf H}]\zeta
\end{align}
as a function of $\alpha$ and $\beta$.
This yields the $\lambda_{max}$ as a point on the real axis since we check the real part of the exponent, and ${\bf G}$ has only the real eigenvalues in our case. The sign of $\lambda_{max}$ will reveal the stability of the eigenmodes and, hence we have the MSF, which is used for delineating the CS boundaries (blue dashed lines) of $\varepsilon$ as shown  in Fig.~\ref{fig2} where $\lambda_{max}$ is negative.
\section{Phase diagram: Ring of  Josephson junctions}
We draw a phase diagram  to identify a variety of collective states including the chimera patterns in a $\alpha-\varepsilon$ parameter plane in Fig.~\ref{fig2} for the three junctions in the ring. All the junctions are in out-of-phase for attractive coupling ($\varepsilon>0$) when the junction voltages evolve in $2\pi$/3 cyclic motion, however, become completely synchronized (CS) in periodic state for a range of repulsive coupling ($\varepsilon<0$). The boundaries of $\varepsilon$ for CS are demarcated at two ends (by two blue dashed lines), which are indicated by a transition of $\lambda_{max}$ to a negative value, and perfectly matching with the numerically simulated boundaries of CS (region A, yellow) in the phase diagram. CS breaks down for both attractive  coupling  at one end (red region), and larger repulsive coupling at the other end (regions B and G). CS transits to regions B (red, asynchronous state in periodic motion) and G (green, mixed state) for larger repulsive coupling. Chimera pattern coexists with asynchronous oscillation in region G (chaotic or higher periodic rotational motion). The ring of junctions becomes chaotic in region C (orange) and remains asynchronous. They form chaotic chimera pattern with two coherent and one incoherent junctions in region D when all the junctions maintain rotational motion. Finally, a periodic state re-emerges in region E (cyan) with a chimera pattern. Two junctions becomes coherent in rotational periodic motion that coexists with a single junction in incoherent periodic libration. This chimera patterns flips with a new pattern when two junctions becomes coherent in libration and one in incoherent rotation; this flipping occurs due to sensitivity in initial conditions. Region F represents mixed states of coexisting chimera pattern and asynchronous states.

\begin{figure}[!h]
	\centering
	\includegraphic[scale=0.5]{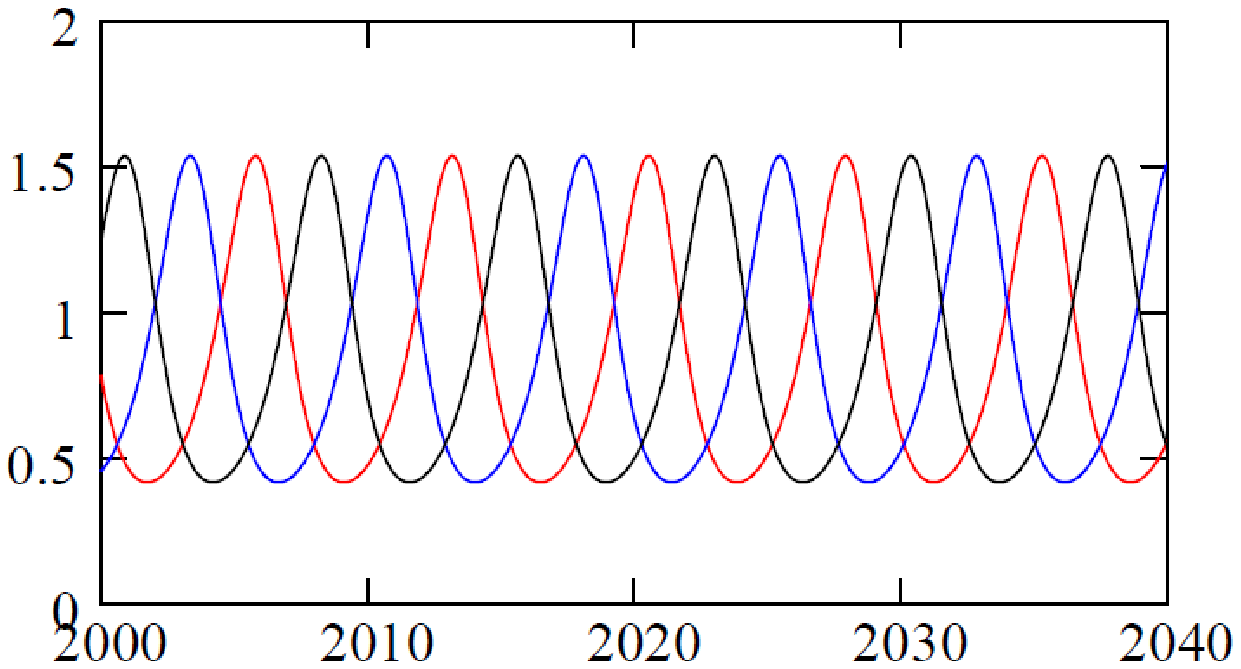}{(a)}{0pt}{0}
	\includegraphic[scale=0.5]{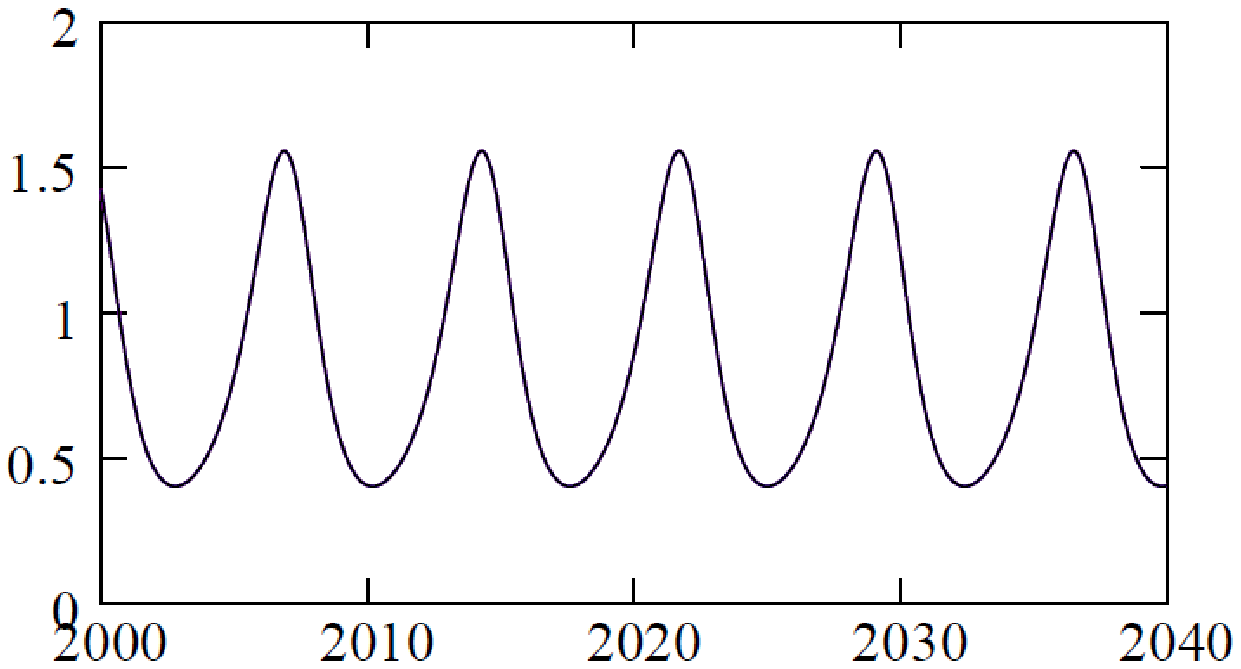}{(b)}{0pt}{0}
	\caption{Time evolution of junction voltages. Voltages $v_i$ ($i=1, 2,3$) in three junctions evolve in (a) out-of-phase $2\pi/3$ motion in cyclic order in time (red, black and blue lines) for $\varepsilon=0.02$, (b) coverge into complete synchrony of amplitude and phase (black line) for $\varepsilon=-0.01$.  Parameters $\alpha=1.45$ and $i=1.5$. }
	\label{fig3}
\end{figure}
\begin{figure*}[!h]
	\centering
	\includegraphic[scale=0.49]{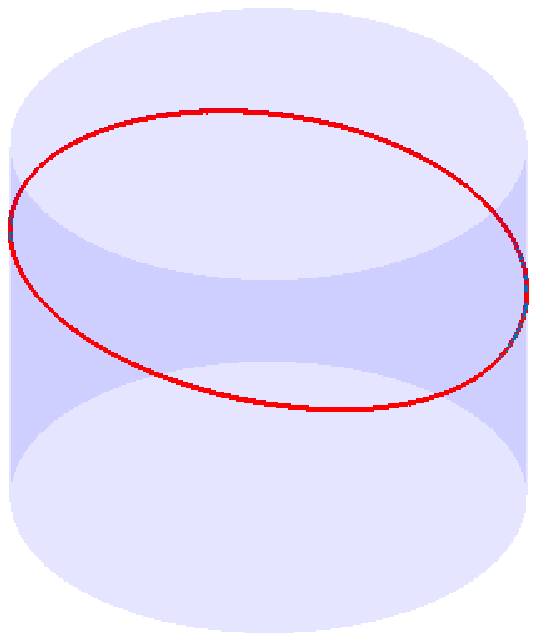}{(a)}{0pt}{0}
	\includegraphic[height=3.5cm, width=6.5cm		]{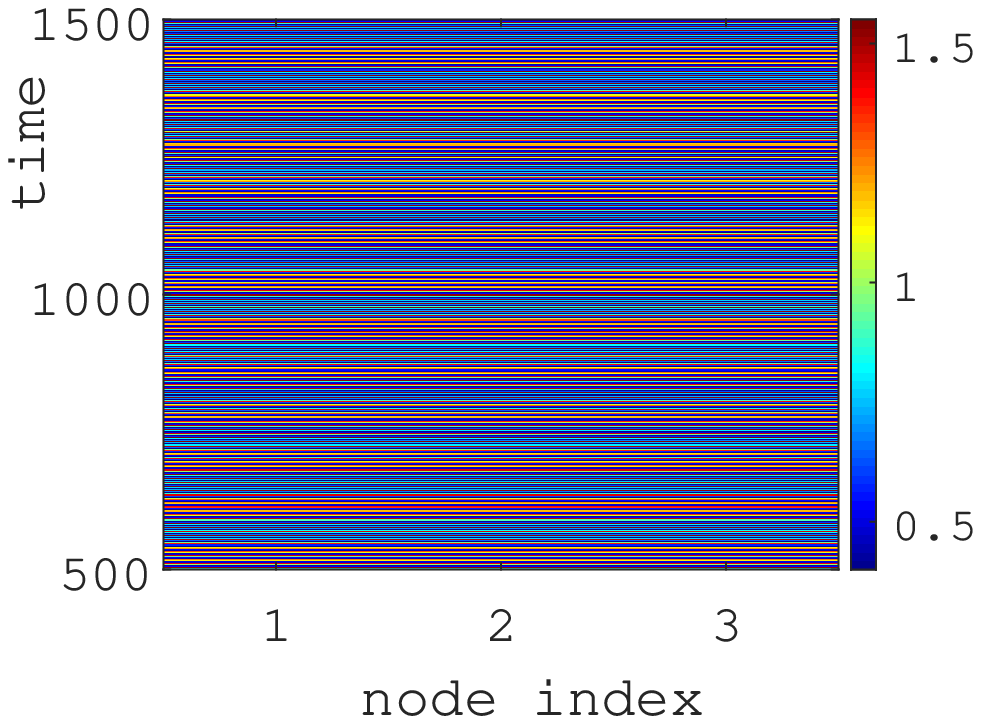}{(b)}{0pt}{0}\\
	\includegraphic[scale=0.49]{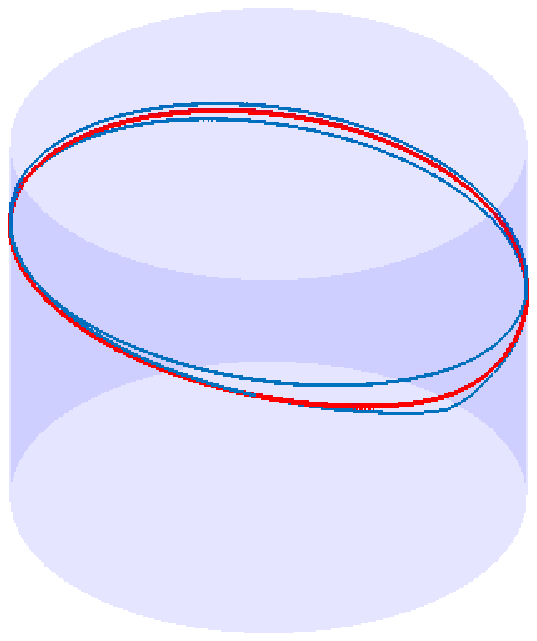}{(c)}{0pt}{0}
	\includegraphic[height=3.5cm, width=6.5cm]{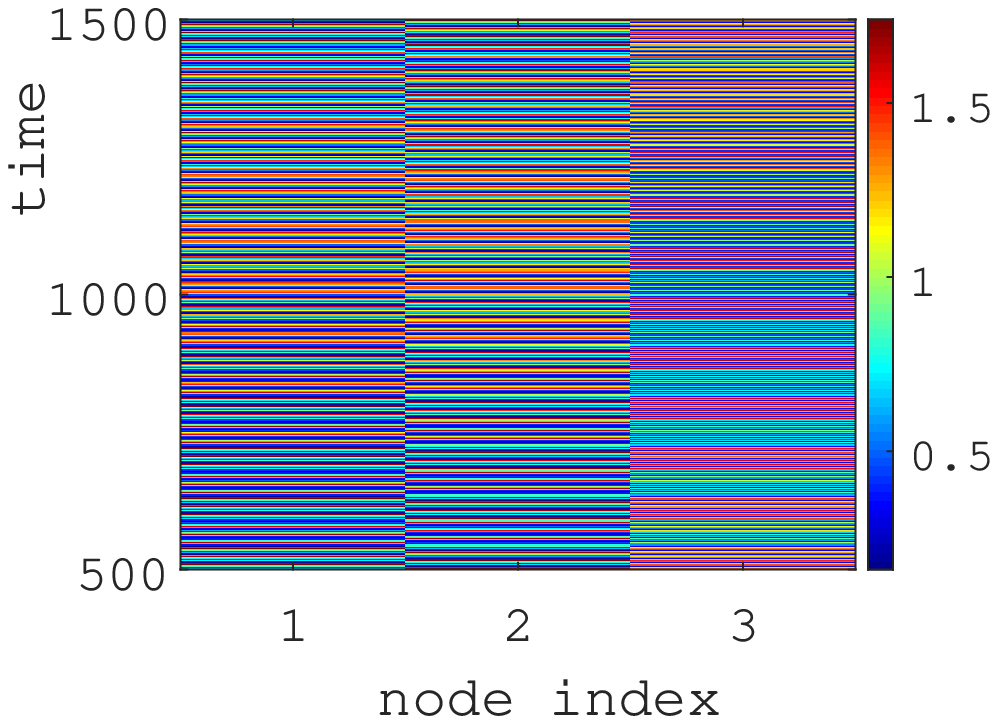}{(d)}{0pt}{0}\\
	\includegraphic[scale=0.49]{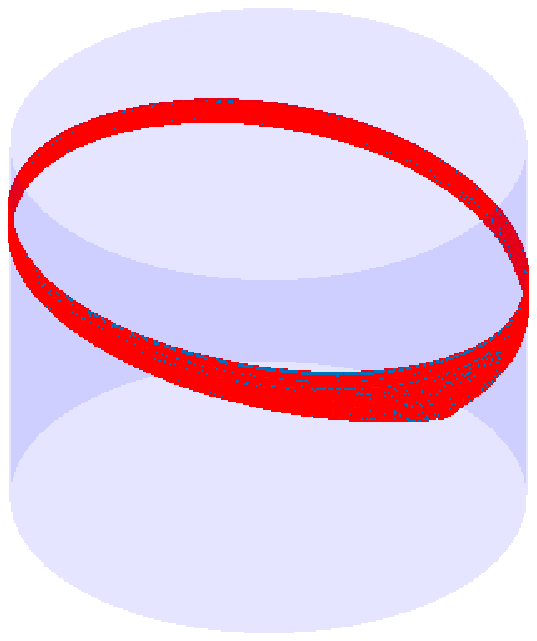}{(e)}{0pt}{0}
	\includegraphic[height=3.5cm, width=6.5cm]{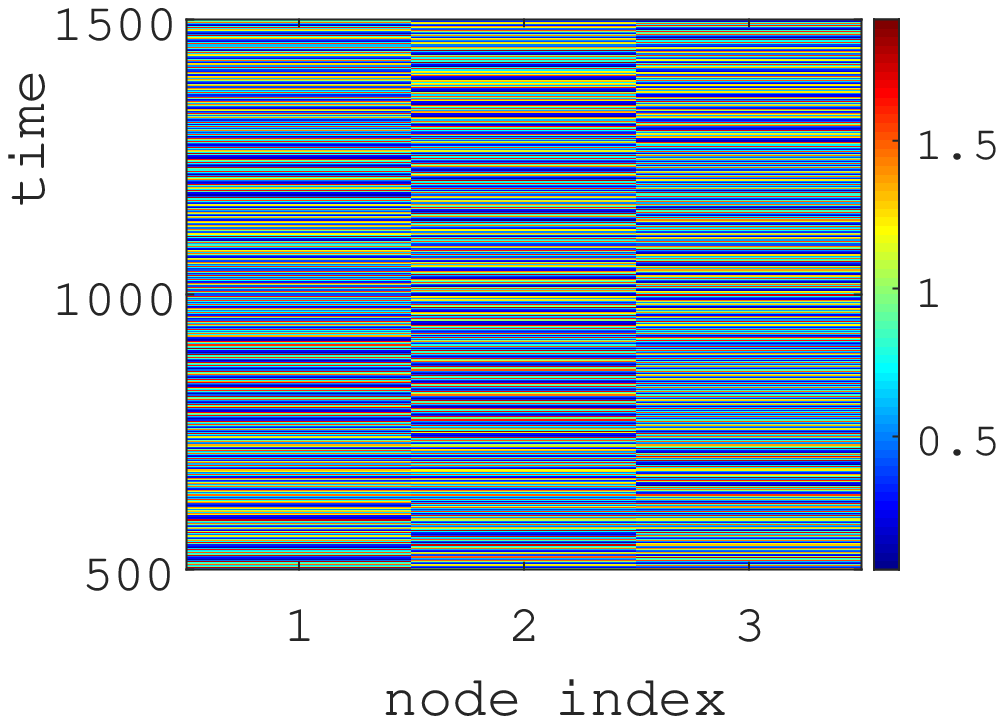}{(f)}{0pt}{0}\\
	\includegraphic[scale=0.49]{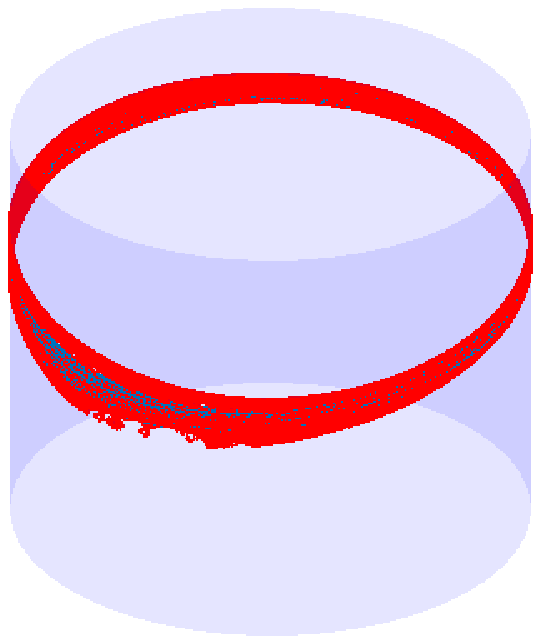}{(g)}{0pt}{0}
	\includegraphic[height=3.5cm, width=6.5cm]{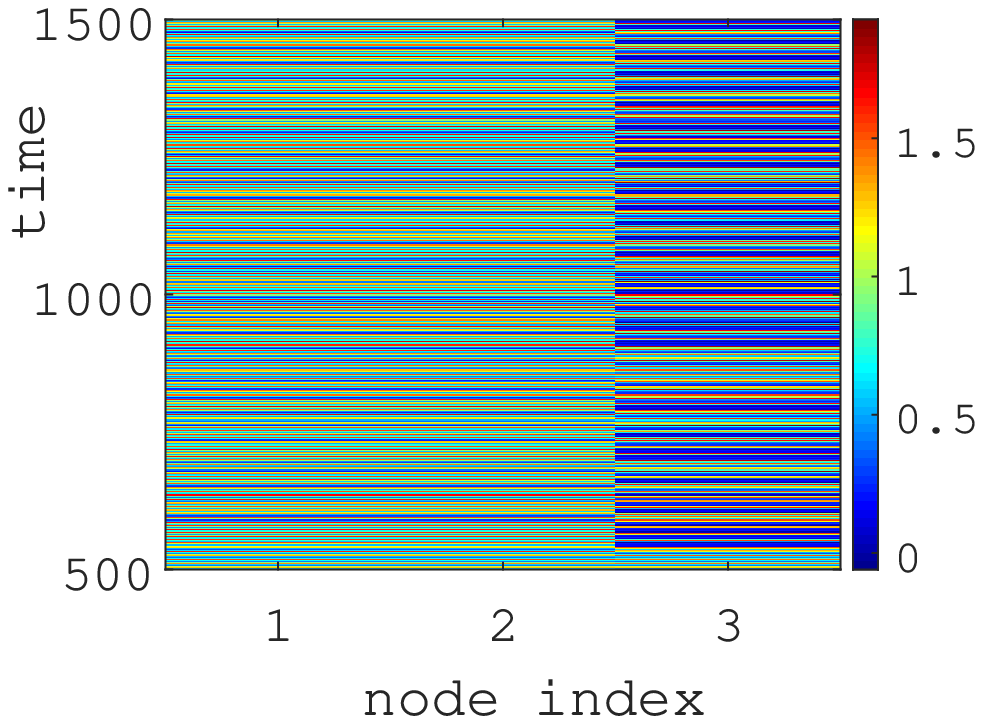}{(h)}{0pt}{0}\\
	\includegraphic[scale=0.49]{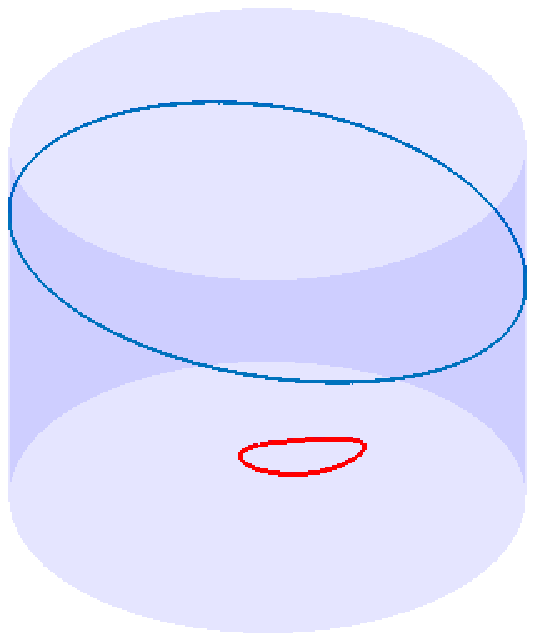}{(i)}{0pt}{0}
	\includegraphic[height=3.5cm, width=6.5cm]{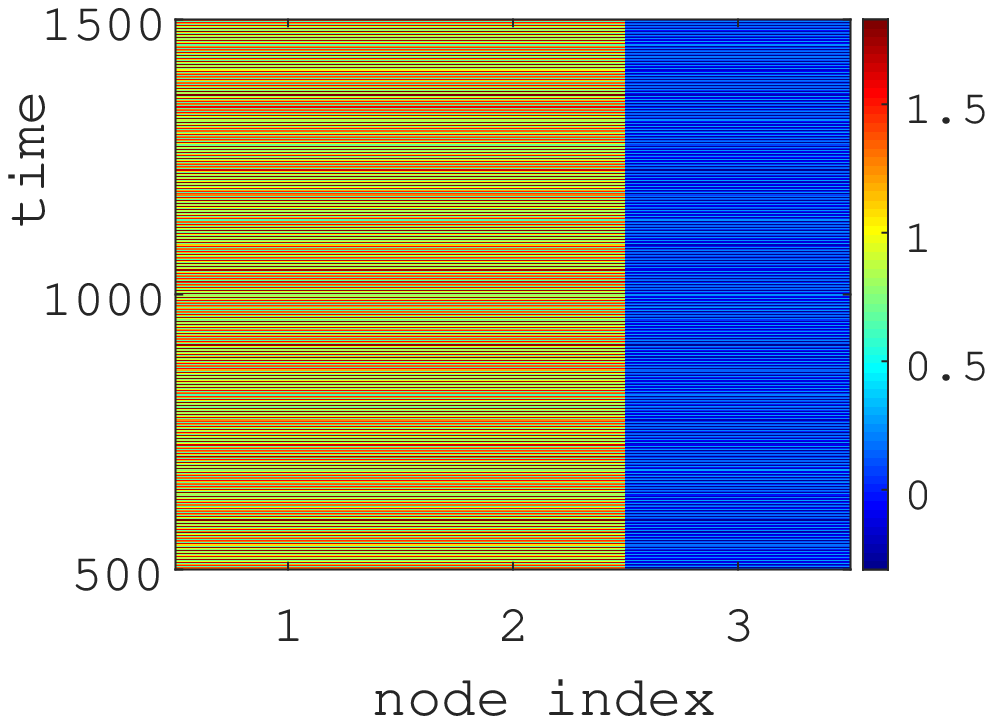}{(j)}{0pt}{0}
	\caption{Smallest Chimeras in a ring of three Josephson junctions.  (a, c, e, g, i) Trajectory of junctions in  $\theta$-$v$ cylindrical plane and (b, d, f, h, j) spatiotemporal evolution of junction voltage $v_i  (i=1, 2, 3)$ in three junctions. Region A: (a) One coherent periodic trajectory (red) is seen with  coherent evolution (b) of all three junctions in time for $\alpha=1.45, \varepsilon=-0.02$. Region B: Three distinctly separate periodic rotational trajectories (c) are seen in incoherent motion (d) in time for $\alpha=1.45, \varepsilon=-0.23$. Region C: Three junctions are in chaotic rotational motion (e) with incoherent spatial evolution (f) in time for $\alpha=1.45, \varepsilon=-0.24$. Region D: Chimera pattern in chaotic motion (g) when two junctions $(1,2)$ are evolving (h) in coherent motion and one ($3$) is incoherent for $\alpha=1.45, \varepsilon=-0.25$. Region E: Chimera pattern in dual motion, (i) one coherent pair of junctions $(1,2)$ in rotational motion  and one junction ($3$) in incoherent librational motion for $\alpha=1.45, \varepsilon=-0.265$.  Regions A-E are identified here as depicted in Fig.~\ref{fig2}. Color bars depict instantaneous values of $v_i$.}
	\label{fig4}
\end{figure*}

\section{Smallest chimeras in three junctions}
Three identical junctions in the ring evolve with $2\pi/3$ phase lag incyclic order for attractive coupling ($\epsilon>0$) and they evolve into a state of complete identical phase and amplitude for a repulsive coupling ($\epsilon<0$) above a threshold as decided by the MSF. The time evolution of the three junctions are plotted in Figs.~\ref{fig3}(a) and \ref{fig3}(b) when the junctions emerge in an out-of-phase state and CS  state, for $\varepsilon=0.02$ and $\varepsilon=-0.01$, respectively. Both the out-of-phase and CS states are true for $N>3$ number of junctions in the ring (results not shown here, but checked for $N=4, 7$). 
We have already recognized two reasonably broad parameter regions of chimera patterns in parameter plane (D and E). Now we demonstrate their exemplary dynamics using the trajectories of the three junctions in a $\theta-v$ cyclindrical plane and their corresponding spatio-temporal dynamics of junction voltage $v_i$ and how they evolve against varying repulsive coupling $\varepsilon$ at a fixed $\alpha=1.45$. Figures \ref{fig4}(a)-(b) represent CS state as confirmed by a plot of the trajectories (red) of three junctions in a $\theta-v$ cylindrical plane (left panel) and its corresponding spatio-temporal plot at right panel. For stronger repulsive interaction, Figs.~\ref{fig4}(c)-(d) show that CS breaks down although the junctions maintains rotational periodic motion, however,  their trajectories are distinctly different (blue, red and green) which is confirmed by their incoherent motion in right panel. The junctions become chaotic for larger repulsive coupling, yet continues with rotational incoherent motion as depicted by their trajectories and spatio-temporal evolution in Figs.~\ref{fig4}(e)-(f). The first sign of chimera pattern appears for further increase in repulsive coupling when all the junctions are in chaotic rotational motion as shown in Fig.~\ref{fig4}(g).  However, two of the junctions (indices 2 and 3) are evolving in coherence of amplitude and phase while the third junction (index 1) is incoherent in rotational motion as depicted in the spatiotemporal plot in Fig.~\ref{fig4}(h). At this stage, the collective dynamics of three junctions switches to periodic motion for a further increase in repulsive coupling. Interestingly, the junctions split into two groups when a pair of junctions remains in coherent rotation (blue) as shown in Fig.~\ref{fig4}(i) and one junction (3) switches to librational motion (red) and incoherent to the other two junctions (1 and 2). This coexisting state of coherence and incoherence in two groups is typically defined as a chimera pattern and confirmed by their spatiotemporal evolution in Fig.~\ref{fig4}(j).

\section{Conclusions} Chimera patterns indeed emerged in a smallest group of three junctions interacting repulsively in a ring. Surprisingly, three junctions in rotational motion when arranged in a ring, emerge into complete synchrony of amplitude and phase for repulsive coupling. For attractive coupling, they maintain an out-of-phase in $2\pi/3$ alternate motion in cyclic order. We have checked the stability of the CS state for a range of repulsive coupling using the MSF for a range of repulsive coupling and it is independent of the  number of junctions in the ring.  We are still working to find an explanation for such a counter-intuitive collective dynamics, however, it is not the target of our present report. 
\par Chimera patterns emerge in three junctions in a ring beyond the CS state when it breaks down for larger repulsive coupling. We define chimera states here as two coherent junctions and one incoherent junction to the other two that  follow the standard definition of chimera states in the literature. We identified two different patterns of chimera states. Three junctions may remain in chaotic rotational motion yet two of them evolve in coherence while the third junction also evolves in chaotic rotational motion, but incoherent to the other two junctions. For larger repulsive coupling, two junctions  evolve coherently in rotational periodic motion while the third junction switches to periodic librational motion and becomes incoherent to the other two junctions. This chimera pattern may flip to another chimera pattern with two coherent junctions in libration and one incoherent junction in rotation with a change in initial conditions.

\noindent {\bf Data availability}: All data available from the authors free on request.


\begin{thebibliography}{100}%
	\makeatletter
	\providecommand \@ifxundefined [1]{%
		\@ifx{#1\undefined}
	}%
	\providecommand \@ifnum [1]{%
		\ifnum #1\expandafter \@firstoftwo
		\else \expandafter \@secondoftwo
		\fi
	}%
	\providecommand \@ifx [1]{%
		\ifx #1\expandafter \@firstoftwo
		\else \expandafter \@secondoftwo
		\fi
	}%
	\providecommand \natexlab [1]{#1}%
	\providecommand \enquote  [1]{``#1''}%
	\providecommand \bibnamefont  [1]{#1}%
	\providecommand \bibfnamefont [1]{#1}%
	\providecommand \citenamefont [1]{#1}%
	\providecommand \href@noop [0]{\@secondoftwo}%
	\providecommand \href [0]{\begingroup \@sanitize@url \@href}%
	\providecommand \@href[1]{\@@startlink{#1}\@@href}%
	\providecommand \@@href[1]{\endgroup#1\@@endlink}%
	\providecommand \@sanitize@url [0]{\catcode `\\12\catcode `\$12\catcode
		`\&12\catcode `\#12\catcode `\^12\catcode `\_12\catcode `\%12\relax}%
	\providecommand \@@startlink[1]{}%
	\providecommand \@@endlink[0]{}%
	\providecommand \url  [0]{\begingroup\@sanitize@url \@url }%
	\providecommand \@url [1]{\endgroup\@href {#1}{\urlprefix }}%
	\providecommand \urlprefix  [0]{URL }%
	\providecommand \Eprint [0]{\href }%
	\providecommand \doibase [0]{https://doi.org/}%
	\providecommand \selectlanguage [0]{\@gobble}%
	\providecommand \bibinfo  [0]{\@secondoftwo}%
	\providecommand \bibfield  [0]{\@secondoftwo}%
	\providecommand \translation [1]{[#1]}%
	\providecommand \BibitemOpen [0]{}%
	\providecommand \bibitemStop [0]{}%
	\providecommand \bibitemNoStop [0]{.\EOS\space}%
	\providecommand \EOS [0]{\spacefactor3000\relax}%
	\providecommand \BibitemShut  [1]{\csname bibitem#1\endcsname}%
	\let\auto@bib@innerbib\@empty
	\bibitem [{\citenamefont {Parastesh}\ \emph {et~al.}(2020)\citenamefont
		{Parastesh}, \citenamefont {Jafari}, \citenamefont {Azarnoush}, \citenamefont
		{Shahriari}, \citenamefont {Wang}, \citenamefont {Boccaletti},\ and\
		\citenamefont {Perc}}]{parastesh2020chimeras}%
	\BibitemOpen
	\bibfield  {author} {\bibinfo {author} {\bibfnamefont {F.}~\bibnamefont
			{Parastesh}}, \bibinfo {author} {\bibfnamefont {S.}~\bibnamefont {Jafari}},
		\bibinfo {author} {\bibfnamefont {H.}~\bibnamefont {Azarnoush}}, \bibinfo
		{author} {\bibfnamefont {Z.}~\bibnamefont {Shahriari}}, \bibinfo {author}
		{\bibfnamefont {Z.}~\bibnamefont {Wang}}, \bibinfo {author} {\bibfnamefont
			{S.}~\bibnamefont {Boccaletti}},\ and\ \bibinfo {author} {\bibfnamefont
			{M.}~\bibnamefont {Perc}},\ }\bibfield  {title} {\bibinfo {title}
		{Chimeras},\ }\href@noop {} {\bibfield  {journal} {\bibinfo  {journal}
			{Physics Reports}\ } (\bibinfo {year} {2020})}\BibitemShut {NoStop}%
	\bibitem [{\citenamefont {Majhi}\ \emph {et~al.}(2019)\citenamefont {Majhi},
		\citenamefont {Bera}, \citenamefont {Ghosh},\ and\ \citenamefont
		{Perc}}]{majhi2019chimera}%
	\BibitemOpen
	\bibfield  {author} {\bibinfo {author} {\bibfnamefont {S.}~\bibnamefont
			{Majhi}}, \bibinfo {author} {\bibfnamefont {B.~K.}\ \bibnamefont {Bera}},
		\bibinfo {author} {\bibfnamefont {D.}~\bibnamefont {Ghosh}},\ and\ \bibinfo
		{author} {\bibfnamefont {M.}~\bibnamefont {Perc}},\ }\bibfield  {title}
	{\bibinfo {title} {Chimera states in neuronal networks: a review},\
	}\href@noop {} {\bibfield  {journal} {\bibinfo  {journal} {Physics of life
				reviews}\ }\textbf {\bibinfo {volume} {28}},\ \bibinfo {pages} {100}
		(\bibinfo {year} {2019})}\BibitemShut {NoStop}%
	\bibitem [{\citenamefont {Wang}\ and\ \citenamefont
		{Liu}(2020)}]{wang2020brief}%
	\BibitemOpen
	\bibfield  {author} {\bibinfo {author} {\bibfnamefont {Z.}~\bibnamefont
			{Wang}}\ and\ \bibinfo {author} {\bibfnamefont {Z.}~\bibnamefont {Liu}},\
	}\bibfield  {title} {\bibinfo {title} {A brief review of chimera state in
			empirical brain networks},\ }\href@noop {} {\bibfield  {journal} {\bibinfo
			{journal} {Frontiers in Physiology}\ }\textbf {\bibinfo {volume} {11}}
		(\bibinfo {year} {2020})}\BibitemShut {NoStop}%
	\bibitem [{\citenamefont {Abrams}\ and\ \citenamefont
		{Strogatz}(2004)}]{abrams2004chimera}%
	\BibitemOpen
	\bibfield  {author} {\bibinfo {author} {\bibfnamefont {D.~M.}\ \bibnamefont
			{Abrams}}\ and\ \bibinfo {author} {\bibfnamefont {S.~H.}\ \bibnamefont
			{Strogatz}},\ }\bibfield  {title} {\bibinfo {title} {Chimera states for
			coupled oscillators},\ }\href@noop {} {\bibfield  {journal} {\bibinfo
			{journal} {Physical review letters}\ }\textbf {\bibinfo {volume} {93}},\
		\bibinfo {pages} {174102} (\bibinfo {year} {2004})}\BibitemShut {NoStop}%
	\bibitem [{\citenamefont {Sethia}\ \emph {et~al.}(2008)\citenamefont {Sethia},
		\citenamefont {Sen},\ and\ \citenamefont {Atay}}]{sethia2008clustered}%
	\BibitemOpen
	\bibfield  {author} {\bibinfo {author} {\bibfnamefont {G.~C.}\ \bibnamefont
			{Sethia}}, \bibinfo {author} {\bibfnamefont {A.}~\bibnamefont {Sen}},\ and\
		\bibinfo {author} {\bibfnamefont {F.~M.}\ \bibnamefont {Atay}},\ }\bibfield
	{title} {\bibinfo {title} {Clustered chimera states in delay-coupled
			oscillator systems},\ }\href@noop {} {\bibfield  {journal} {\bibinfo
			{journal} {Physical review letters}\ }\textbf {\bibinfo {volume} {100}},\
		\bibinfo {pages} {144102} (\bibinfo {year} {2008})}\BibitemShut {NoStop}%
	\bibitem [{\citenamefont {Martens}\ \emph {et~al.}(2013)\citenamefont
		{Martens}, \citenamefont {Thutupalli}, \citenamefont {Fourriere},\ and\
		\citenamefont {Hallatschek}}]{martens2013chimera}%
	\BibitemOpen
	\bibfield  {author} {\bibinfo {author} {\bibfnamefont {E.~A.}\ \bibnamefont
			{Martens}}, \bibinfo {author} {\bibfnamefont {S.}~\bibnamefont {Thutupalli}},
		\bibinfo {author} {\bibfnamefont {A.}~\bibnamefont {Fourriere}},\ and\
		\bibinfo {author} {\bibfnamefont {O.}~\bibnamefont {Hallatschek}},\
	}\bibfield  {title} {\bibinfo {title} {Chimera states in mechanical
			oscillator networks},\ }\href@noop {} {\bibfield  {journal} {\bibinfo
			{journal} {Proceedings of the National Academy of Sciences}\ }\textbf
		{\bibinfo {volume} {110}},\ \bibinfo {pages} {10563} (\bibinfo {year}
		{2013})}\BibitemShut {NoStop}%
	\bibitem [{\citenamefont {Laing}(2009)}]{laing2009chimera}%
	\BibitemOpen
	\bibfield  {author} {\bibinfo {author} {\bibfnamefont {C.~R.}\ \bibnamefont
			{Laing}},\ }\bibfield  {title} {\bibinfo {title} {Chimera states in
			heterogeneous networks},\ }\href@noop {} {\bibfield  {journal} {\bibinfo
			{journal} {Chaos: An Interdisciplinary Journal of Nonlinear Science}\
		}\textbf {\bibinfo {volume} {19}},\ \bibinfo {pages} {013113} (\bibinfo
		{year} {2009})}\BibitemShut {NoStop}%
	\bibitem [{\citenamefont {Gopal}\ \emph {et~al.}(2014)\citenamefont {Gopal},
		\citenamefont {Chandrasekar}, \citenamefont {Venkatesan},\ and\ \citenamefont
		{Lakshmanan}}]{gopal2014observation}%
	\BibitemOpen
	\bibfield  {author} {\bibinfo {author} {\bibfnamefont {R.}~\bibnamefont
			{Gopal}}, \bibinfo {author} {\bibfnamefont {V.}~\bibnamefont {Chandrasekar}},
		\bibinfo {author} {\bibfnamefont {A.}~\bibnamefont {Venkatesan}},\ and\
		\bibinfo {author} {\bibfnamefont {M.}~\bibnamefont {Lakshmanan}},\ }\bibfield
	{title} {\bibinfo {title} {Observation and characterization of chimera
			states in coupled dynamical systems with nonlocal coupling},\ }\href@noop {}
	{\bibfield  {journal} {\bibinfo  {journal} {Physical review E}\ }\textbf
		{\bibinfo {volume} {89}},\ \bibinfo {pages} {052914} (\bibinfo {year}
		{2014})}\BibitemShut {NoStop}%
	\bibitem [{\citenamefont {Hagerstrom}\ \emph {et~al.}(2012)\citenamefont
		{Hagerstrom}, \citenamefont {Murphy}, \citenamefont {Roy}, \citenamefont
		{H{\"o}vel}, \citenamefont {Omelchenko},\ and\ \citenamefont
		{Sch{\"o}ll}}]{hagerstrom2012experimental}%
	\BibitemOpen
	\bibfield  {author} {\bibinfo {author} {\bibfnamefont {A.~M.}\ \bibnamefont
			{Hagerstrom}}, \bibinfo {author} {\bibfnamefont {T.~E.}\ \bibnamefont
			{Murphy}}, \bibinfo {author} {\bibfnamefont {R.}~\bibnamefont {Roy}},
		\bibinfo {author} {\bibfnamefont {P.}~\bibnamefont {H{\"o}vel}}, \bibinfo
		{author} {\bibfnamefont {I.}~\bibnamefont {Omelchenko}},\ and\ \bibinfo
		{author} {\bibfnamefont {E.}~\bibnamefont {Sch{\"o}ll}},\ }\bibfield  {title}
	{\bibinfo {title} {Experimental observation of chimeras in coupled-map
			lattices},\ }\href@noop {} {\bibfield  {journal} {\bibinfo  {journal} {Nature
				Physics}\ }\textbf {\bibinfo {volume} {8}},\ \bibinfo {pages} {658} (\bibinfo
		{year} {2012})}\BibitemShut {NoStop}%
	\bibitem [{\citenamefont {Hart}\ \emph {et~al.}(2019)\citenamefont {Hart},
		\citenamefont {Larger}, \citenamefont {Murphy},\ and\ \citenamefont
		{Roy}}]{hart2019delayed}%
	\BibitemOpen
	\bibfield  {author} {\bibinfo {author} {\bibfnamefont {J.~D.}\ \bibnamefont
			{Hart}}, \bibinfo {author} {\bibfnamefont {L.}~\bibnamefont {Larger}},
		\bibinfo {author} {\bibfnamefont {T.~E.}\ \bibnamefont {Murphy}},\ and\
		\bibinfo {author} {\bibfnamefont {R.}~\bibnamefont {Roy}},\ }\bibfield
	{title} {\bibinfo {title} {Delayed dynamical systems: Networks, chimeras and
			reservoir computing},\ }\href@noop {} {\bibfield  {journal} {\bibinfo
			{journal} {Philosophical Transactions of the Royal Society A}\ }\textbf
		{\bibinfo {volume} {377}},\ \bibinfo {pages} {20180123} (\bibinfo {year}
		{2019})}\BibitemShut {NoStop}%
	\bibitem [{\citenamefont {Omelchenko}\ \emph {et~al.}(2013)\citenamefont
		{Omelchenko}, \citenamefont {Omelâ€™chenko}, \citenamefont {H{\"o}vel},\ and\
		\citenamefont {Sch{\"o}ll}}]{omelchenko2013nonlocal}%
	\BibitemOpen
	\bibfield  {author} {\bibinfo {author} {\bibfnamefont {I.}~\bibnamefont
			{Omelchenko}}, \bibinfo {author} {\bibfnamefont {E.}~\bibnamefont
			{Omelâ€™chenko}}, \bibinfo {author} {\bibfnamefont {P.}~\bibnamefont
			{H{\"o}vel}},\ and\ \bibinfo {author} {\bibfnamefont {E.}~\bibnamefont
			{Sch{\"o}ll}},\ }\bibfield  {title} {\bibinfo {title} {When nonlocal coupling
			between oscillators becomes stronger: patched synchrony or multichimera
			states},\ }\href@noop {} {\bibfield  {journal} {\bibinfo  {journal} {Physical
				review letters}\ }\textbf {\bibinfo {volume} {110}},\ \bibinfo {pages}
		{224101} (\bibinfo {year} {2013})}\BibitemShut {NoStop}%
	\bibitem [{\citenamefont {Kuramoto}\ and\ \citenamefont
		{Battogtokh}(2002)}]{kuramoto2002coexistence}%
	\BibitemOpen
	\bibfield  {author} {\bibinfo {author} {\bibfnamefont {Y.}~\bibnamefont
			{Kuramoto}}\ and\ \bibinfo {author} {\bibfnamefont {D.}~\bibnamefont
			{Battogtokh}},\ }\bibfield  {title} {\bibinfo {title} {Coexistence of
			coherence and incoherence in nonlocally coupled phase oscillators},\
	}\href@noop {} {\bibfield  {journal} {\bibinfo  {journal} {arXiv preprint
				cond-mat/0210694}\ } (\bibinfo {year} {2002})}\BibitemShut {NoStop}%
	\bibitem [{\citenamefont {Wolfrum}\ and\ \citenamefont
		{Omelâ€™chenko}(2011)}]{wolfrum2011chimera}%
	\BibitemOpen
	\bibfield  {author} {\bibinfo {author} {\bibfnamefont {M.}~\bibnamefont
			{Wolfrum}}\ and\ \bibinfo {author} {\bibfnamefont {E.}~\bibnamefont
			{Omelâ€™chenko}},\ }\bibfield  {title} {\bibinfo {title} {Chimera states are
			chaotic transients},\ }\href@noop {} {\bibfield  {journal} {\bibinfo
			{journal} {Physical Review E}\ }\textbf {\bibinfo {volume} {84}},\ \bibinfo
		{pages} {015201} (\bibinfo {year} {2011})}\BibitemShut {NoStop}%
	\bibitem [{\citenamefont {Rosin}\ \emph {et~al.}(2014)\citenamefont {Rosin},
		\citenamefont {Rontani}, \citenamefont {Haynes}, \citenamefont {Sch{\"o}ll},\
		and\ \citenamefont {Gauthier}}]{rosin2014transient}%
	\BibitemOpen
	\bibfield  {author} {\bibinfo {author} {\bibfnamefont {D.~P.}\ \bibnamefont
			{Rosin}}, \bibinfo {author} {\bibfnamefont {D.}~\bibnamefont {Rontani}},
		\bibinfo {author} {\bibfnamefont {N.~D.}\ \bibnamefont {Haynes}}, \bibinfo
		{author} {\bibfnamefont {E.}~\bibnamefont {Sch{\"o}ll}},\ and\ \bibinfo
		{author} {\bibfnamefont {D.~J.}\ \bibnamefont {Gauthier}},\ }\bibfield
	{title} {\bibinfo {title} {Transient scaling and resurgence of chimera states
			in networks of boolean phase oscillators},\ }\href@noop {} {\bibfield
		{journal} {\bibinfo  {journal} {Physical Review E}\ }\textbf {\bibinfo
			{volume} {90}},\ \bibinfo {pages} {030902} (\bibinfo {year}
		{2014})}\BibitemShut {NoStop}%
	\bibitem [{\citenamefont {Omelâ€™chenko}(2018)}]{omel2018mathematics}%
	\BibitemOpen
	\bibfield  {author} {\bibinfo {author} {\bibfnamefont {O.~E.}\ \bibnamefont
			{Omelâ€™chenko}},\ }\bibfield  {title} {\bibinfo {title} {The mathematics
			behind chimera states},\ }\href@noop {} {\bibfield  {journal} {\bibinfo
			{journal} {Nonlinearity}\ }\textbf {\bibinfo {volume} {31}},\ \bibinfo
		{pages} {R121} (\bibinfo {year} {2018})}\BibitemShut {NoStop}%
	\bibitem [{\citenamefont {Laing}(2019)}]{laing2019dynamics}%
	\BibitemOpen
	\bibfield  {author} {\bibinfo {author} {\bibfnamefont {C.~R.}\ \bibnamefont
			{Laing}},\ }\bibfield  {title} {\bibinfo {title} {Dynamics and stability of
			chimera states in two coupled populations of oscillators},\ }\href@noop {}
	{\bibfield  {journal} {\bibinfo  {journal} {Physical Review E}\ }\textbf
		{\bibinfo {volume} {100}},\ \bibinfo {pages} {042211} (\bibinfo {year}
		{2019})}\BibitemShut {NoStop}%
	\bibitem [{\citenamefont {Pecora}\ \emph {et~al.}(2014)\citenamefont {Pecora},
		\citenamefont {Sorrentino}, \citenamefont {Hagerstrom}, \citenamefont
		{Murphy},\ and\ \citenamefont {Roy}}]{pecora2014cluster}%
	\BibitemOpen
	\bibfield  {author} {\bibinfo {author} {\bibfnamefont {L.~M.}\ \bibnamefont
			{Pecora}}, \bibinfo {author} {\bibfnamefont {F.}~\bibnamefont {Sorrentino}},
		\bibinfo {author} {\bibfnamefont {A.~M.}\ \bibnamefont {Hagerstrom}},
		\bibinfo {author} {\bibfnamefont {T.~E.}\ \bibnamefont {Murphy}},\ and\
		\bibinfo {author} {\bibfnamefont {R.}~\bibnamefont {Roy}},\ }\bibfield
	{title} {\bibinfo {title} {Cluster synchronization and isolated
			desynchronization in complex networks with symmetries},\ }\href@noop {}
	{\bibfield  {journal} {\bibinfo  {journal} {Nature communications}\ }\textbf
		{\bibinfo {volume} {5}},\ \bibinfo {pages} {1} (\bibinfo {year}
		{2014})}\BibitemShut {NoStop}%
	\bibitem [{\citenamefont {Maksimenko}\ \emph {et~al.}(2016)\citenamefont
		{Maksimenko}, \citenamefont {Makarov}, \citenamefont {Bera}, \citenamefont
		{Ghosh}, \citenamefont {Dana}, \citenamefont {Goremyko}, \citenamefont
		{Frolov}, \citenamefont {Koronovskii},\ and\ \citenamefont
		{Hramov}}]{maksimenko2016excitation}%
	\BibitemOpen
	\bibfield  {author} {\bibinfo {author} {\bibfnamefont {V.~A.}\ \bibnamefont
			{Maksimenko}}, \bibinfo {author} {\bibfnamefont {V.~V.}\ \bibnamefont
			{Makarov}}, \bibinfo {author} {\bibfnamefont {B.~K.}\ \bibnamefont {Bera}},
		\bibinfo {author} {\bibfnamefont {D.}~\bibnamefont {Ghosh}}, \bibinfo
		{author} {\bibfnamefont {S.~K.}\ \bibnamefont {Dana}}, \bibinfo {author}
		{\bibfnamefont {M.~V.}\ \bibnamefont {Goremyko}}, \bibinfo {author}
		{\bibfnamefont {N.~S.}\ \bibnamefont {Frolov}}, \bibinfo {author}
		{\bibfnamefont {A.~A.}\ \bibnamefont {Koronovskii}},\ and\ \bibinfo {author}
		{\bibfnamefont {A.~E.}\ \bibnamefont {Hramov}},\ }\bibfield  {title}
	{\bibinfo {title} {Excitation and suppression of chimera states by
			multiplexing},\ }\href@noop {} {\bibfield  {journal} {\bibinfo  {journal}
			{Physical Review E}\ }\textbf {\bibinfo {volume} {94}},\ \bibinfo {pages}
		{052205} (\bibinfo {year} {2016})}\BibitemShut {NoStop}%
	\bibitem [{\citenamefont {Ghosh}\ and\ \citenamefont
		{Jalan}(2016)}]{ghosh2016emergence}%
	\BibitemOpen
	\bibfield  {author} {\bibinfo {author} {\bibfnamefont {S.}~\bibnamefont
			{Ghosh}}\ and\ \bibinfo {author} {\bibfnamefont {S.}~\bibnamefont {Jalan}},\
	}\bibfield  {title} {\bibinfo {title} {Emergence of chimera in multiplex
			network},\ }\href@noop {} {\bibfield  {journal} {\bibinfo  {journal}
			{International Journal of Bifurcation and Chaos}\ }\textbf {\bibinfo {volume}
			{26}},\ \bibinfo {pages} {1650120} (\bibinfo {year} {2016})}\BibitemShut
	{NoStop}%
	\bibitem [{\citenamefont {Sawicki}\ \emph {et~al.}(2018)\citenamefont
		{Sawicki}, \citenamefont {Omelchenko}, \citenamefont {Zakharova},\ and\
		\citenamefont {Sch{\"o}ll}}]{sawicki2018synchronization}%
	\BibitemOpen
	\bibfield  {author} {\bibinfo {author} {\bibfnamefont {J.}~\bibnamefont
			{Sawicki}}, \bibinfo {author} {\bibfnamefont {I.}~\bibnamefont {Omelchenko}},
		\bibinfo {author} {\bibfnamefont {A.}~\bibnamefont {Zakharova}},\ and\
		\bibinfo {author} {\bibfnamefont {E.}~\bibnamefont {Sch{\"o}ll}},\ }\bibfield
	{title} {\bibinfo {title} {Synchronization scenarios of chimeras in
			multiplex networks},\ }\href@noop {} {\bibfield  {journal} {\bibinfo
			{journal} {The European Physical Journal Special Topics}\ }\textbf {\bibinfo
			{volume} {227}},\ \bibinfo {pages} {1161} (\bibinfo {year}
		{2018})}\BibitemShut {NoStop}%
	\bibitem [{\citenamefont {Ruzzene}\ \emph {et~al.}(2020)\citenamefont
		{Ruzzene}, \citenamefont {Omelchenko}, \citenamefont {Sawicki}, \citenamefont
		{Zakharova}, \citenamefont {Sch{\"o}ll},\ and\ \citenamefont
		{Andrzejak}}]{ruzzene2020remote}%
	\BibitemOpen
	\bibfield  {author} {\bibinfo {author} {\bibfnamefont {G.}~\bibnamefont
			{Ruzzene}}, \bibinfo {author} {\bibfnamefont {I.}~\bibnamefont {Omelchenko}},
		\bibinfo {author} {\bibfnamefont {J.}~\bibnamefont {Sawicki}}, \bibinfo
		{author} {\bibfnamefont {A.}~\bibnamefont {Zakharova}}, \bibinfo {author}
		{\bibfnamefont {E.}~\bibnamefont {Sch{\"o}ll}},\ and\ \bibinfo {author}
		{\bibfnamefont {R.~G.}\ \bibnamefont {Andrzejak}},\ }\bibfield  {title}
	{\bibinfo {title} {Remote pacemaker control of chimera states in multilayer
			networks of neurons},\ }\href@noop {} {\bibfield  {journal} {\bibinfo
			{journal} {Physical Review E}\ }\textbf {\bibinfo {volume} {102}},\ \bibinfo
		{pages} {052216} (\bibinfo {year} {2020})}\BibitemShut {NoStop}%
	\bibitem [{\citenamefont {Maistrenko}\ \emph {et~al.}(2015)\citenamefont
		{Maistrenko}, \citenamefont {Sudakov}, \citenamefont {Osiv},\ and\
		\citenamefont {Maistrenko}}]{maistrenko2015chimera}%
	\BibitemOpen
	\bibfield  {author} {\bibinfo {author} {\bibfnamefont {Y.}~\bibnamefont
			{Maistrenko}}, \bibinfo {author} {\bibfnamefont {O.}~\bibnamefont {Sudakov}},
		\bibinfo {author} {\bibfnamefont {O.}~\bibnamefont {Osiv}},\ and\ \bibinfo
		{author} {\bibfnamefont {V.}~\bibnamefont {Maistrenko}},\ }\bibfield  {title}
	{\bibinfo {title} {Chimera states in three dimensions},\ }\href@noop {}
	{\bibfield  {journal} {\bibinfo  {journal} {New Journal of Physics}\ }\textbf
		{\bibinfo {volume} {17}},\ \bibinfo {pages} {073037} (\bibinfo {year}
		{2015})}\BibitemShut {NoStop}%
	\bibitem [{\citenamefont {Kundu}\ \emph {et~al.}(2019)\citenamefont {Kundu},
		\citenamefont {Bera}, \citenamefont {Ghosh},\ and\ \citenamefont
		{Lakshmanan}}]{kundu2019chimera}%
	\BibitemOpen
	\bibfield  {author} {\bibinfo {author} {\bibfnamefont {S.}~\bibnamefont
			{Kundu}}, \bibinfo {author} {\bibfnamefont {B.~K.}\ \bibnamefont {Bera}},
		\bibinfo {author} {\bibfnamefont {D.}~\bibnamefont {Ghosh}},\ and\ \bibinfo
		{author} {\bibfnamefont {M.}~\bibnamefont {Lakshmanan}},\ }\bibfield  {title}
	{\bibinfo {title} {Chimera patterns in three-dimensional locally coupled
			systems},\ }\href@noop {} {\bibfield  {journal} {\bibinfo  {journal}
			{Physical Review E}\ }\textbf {\bibinfo {volume} {99}},\ \bibinfo {pages}
		{022204} (\bibinfo {year} {2019})}\BibitemShut {NoStop}%
	\bibitem [{\citenamefont {Kasimatis}\ \emph {et~al.}(2018)\citenamefont
		{Kasimatis}, \citenamefont {Hizanidis},\ and\ \citenamefont
		{Provata}}]{kasimatis2018three}%
	\BibitemOpen
	\bibfield  {author} {\bibinfo {author} {\bibfnamefont {T.}~\bibnamefont
			{Kasimatis}}, \bibinfo {author} {\bibfnamefont {J.}~\bibnamefont
			{Hizanidis}},\ and\ \bibinfo {author} {\bibfnamefont {A.}~\bibnamefont
			{Provata}},\ }\bibfield  {title} {\bibinfo {title} {Three-dimensional chimera
			patterns in networks of spiking neuron oscillators},\ }\href@noop {}
	{\bibfield  {journal} {\bibinfo  {journal} {Physical Review E}\ }\textbf
		{\bibinfo {volume} {97}},\ \bibinfo {pages} {052213} (\bibinfo {year}
		{2018})}\BibitemShut {NoStop}%
	\bibitem [{\citenamefont {Omelâ€™chenko}(2019)}]{omel2019traveling}%
	\BibitemOpen
	\bibfield  {author} {\bibinfo {author} {\bibfnamefont {O.}~\bibnamefont
			{Omelâ€™chenko}},\ }\bibfield  {title} {\bibinfo {title} {Traveling chimera
			states},\ }\href@noop {} {\bibfield  {journal} {\bibinfo  {journal} {Journal
				of Physics A: Mathematical and Theoretical}\ }\textbf {\bibinfo {volume}
			{52}},\ \bibinfo {pages} {104001} (\bibinfo {year} {2019})}\BibitemShut
	{NoStop}%
	\bibitem [{\citenamefont {Dudkowski}\ \emph {et~al.}(2019)\citenamefont
		{Dudkowski}, \citenamefont {Czo{\l}czy{\'n}ski},\ and\ \citenamefont
		{Kapitaniak}}]{dudkowski2019traveling}%
	\BibitemOpen
	\bibfield  {author} {\bibinfo {author} {\bibfnamefont {D.}~\bibnamefont
			{Dudkowski}}, \bibinfo {author} {\bibfnamefont {K.}~\bibnamefont
			{Czo{\l}czy{\'n}ski}},\ and\ \bibinfo {author} {\bibfnamefont
			{T.}~\bibnamefont {Kapitaniak}},\ }\bibfield  {title} {\bibinfo {title}
		{Traveling chimera states for coupled pendula},\ }\href@noop {} {\bibfield
		{journal} {\bibinfo  {journal} {Nonlinear Dynamics}\ }\textbf {\bibinfo
			{volume} {95}},\ \bibinfo {pages} {1859} (\bibinfo {year}
		{2019})}\BibitemShut {NoStop}%
	\bibitem [{\citenamefont {Alvarez-Socorro}\ \emph {et~al.}(2021)\citenamefont
		{Alvarez-Socorro}, \citenamefont {Clerc},\ and\ \citenamefont
		{Verschueren}}]{alvarez2021traveling}%
	\BibitemOpen
	\bibfield  {author} {\bibinfo {author} {\bibfnamefont {A.}~\bibnamefont
			{Alvarez-Socorro}}, \bibinfo {author} {\bibfnamefont {M.~G.}\ \bibnamefont
			{Clerc}},\ and\ \bibinfo {author} {\bibfnamefont {N.}~\bibnamefont
			{Verschueren}},\ }\bibfield  {title} {\bibinfo {title} {Traveling chimera
			states in continuous media},\ }\href@noop {} {\bibfield  {journal} {\bibinfo
			{journal} {Communications in Nonlinear Science and Numerical Simulation}\
		}\textbf {\bibinfo {volume} {94}},\ \bibinfo {pages} {105559} (\bibinfo
		{year} {2021})}\BibitemShut {NoStop}%
	\bibitem [{\citenamefont {Martens}\ \emph {et~al.}(2010)\citenamefont
		{Martens}, \citenamefont {Laing},\ and\ \citenamefont
		{Strogatz}}]{martens2010solvable}%
	\BibitemOpen
	\bibfield  {author} {\bibinfo {author} {\bibfnamefont {E.~A.}\ \bibnamefont
			{Martens}}, \bibinfo {author} {\bibfnamefont {C.~R.}\ \bibnamefont {Laing}},\
		and\ \bibinfo {author} {\bibfnamefont {S.~H.}\ \bibnamefont {Strogatz}},\
	}\bibfield  {title} {\bibinfo {title} {Solvable model of spiral wave
			chimeras},\ }\href@noop {} {\bibfield  {journal} {\bibinfo  {journal}
			{Physical review letters}\ }\textbf {\bibinfo {volume} {104}},\ \bibinfo
		{pages} {044101} (\bibinfo {year} {2010})}\BibitemShut {NoStop}%
	\bibitem [{\citenamefont {Gu}\ \emph {et~al.}(2013)\citenamefont {Gu},
		\citenamefont {St-Yves},\ and\ \citenamefont {Davidsen}}]{gu2013spiral}%
	\BibitemOpen
	\bibfield  {author} {\bibinfo {author} {\bibfnamefont {C.}~\bibnamefont
			{Gu}}, \bibinfo {author} {\bibfnamefont {G.}~\bibnamefont {St-Yves}},\ and\
		\bibinfo {author} {\bibfnamefont {J.}~\bibnamefont {Davidsen}},\ }\bibfield
	{title} {\bibinfo {title} {Spiral wave chimeras in complex oscillatory and
			chaotic systems},\ }\href@noop {} {\bibfield  {journal} {\bibinfo  {journal}
			{Physical review letters}\ }\textbf {\bibinfo {volume} {111}},\ \bibinfo
		{pages} {134101} (\bibinfo {year} {2013})}\BibitemShut {NoStop}%
	\bibitem [{\citenamefont {Bera}\ \emph {et~al.}(2016)\citenamefont {Bera},
		\citenamefont {Ghosh},\ and\ \citenamefont {Lakshmanan}}]{bera2016chimera}%
	\BibitemOpen
	\bibfield  {author} {\bibinfo {author} {\bibfnamefont {B.~K.}\ \bibnamefont
			{Bera}}, \bibinfo {author} {\bibfnamefont {D.}~\bibnamefont {Ghosh}},\ and\
		\bibinfo {author} {\bibfnamefont {M.}~\bibnamefont {Lakshmanan}},\ }\bibfield
	{title} {\bibinfo {title} {Chimera states in bursting neurons},\ }\href@noop
	{} {\bibfield  {journal} {\bibinfo  {journal} {Physical Review E}\ }\textbf
		{\bibinfo {volume} {93}},\ \bibinfo {pages} {012205} (\bibinfo {year}
		{2016})}\BibitemShut {NoStop}%
	\bibitem [{\citenamefont {Banerjee}\ \emph {et~al.}(2016)\citenamefont
		{Banerjee}, \citenamefont {Dutta}, \citenamefont {Zakharova},\ and\
		\citenamefont {Sch{\"o}ll}}]{banerjee2016chimera}%
	\BibitemOpen
	\bibfield  {author} {\bibinfo {author} {\bibfnamefont {T.}~\bibnamefont
			{Banerjee}}, \bibinfo {author} {\bibfnamefont {P.~S.}\ \bibnamefont {Dutta}},
		\bibinfo {author} {\bibfnamefont {A.}~\bibnamefont {Zakharova}},\ and\
		\bibinfo {author} {\bibfnamefont {E.}~\bibnamefont {Sch{\"o}ll}},\ }\bibfield
	{title} {\bibinfo {title} {Chimera patterns induced by distance-dependent
			power-law coupling in ecological networks},\ }\href@noop {} {\bibfield
		{journal} {\bibinfo  {journal} {Physical Review E}\ }\textbf {\bibinfo
			{volume} {94}},\ \bibinfo {pages} {032206} (\bibinfo {year}
		{2016})}\BibitemShut {NoStop}%
	\bibitem [{\citenamefont {Saha}\ \emph {et~al.}(2019)\citenamefont {Saha},
		\citenamefont {Bairagi},\ and\ \citenamefont {Dana}}]{saha2019chimera}%
	\BibitemOpen
	\bibfield  {author} {\bibinfo {author} {\bibfnamefont {S.}~\bibnamefont
			{Saha}}, \bibinfo {author} {\bibfnamefont {N.}~\bibnamefont {Bairagi}},\ and\
		\bibinfo {author} {\bibfnamefont {S.~K.}\ \bibnamefont {Dana}},\ }\bibfield
	{title} {\bibinfo {title} {Chimera states in ecological network under
			weighted mean-field dispersal of species},\ }\href@noop {} {\bibfield
		{journal} {\bibinfo  {journal} {Frontiers in Applied Mathematics and
				Statistics}\ }\textbf {\bibinfo {volume} {5}},\ \bibinfo {pages} {15}
		(\bibinfo {year} {2019})}\BibitemShut {NoStop}%
	\bibitem [{\citenamefont {Hizanidis}\ \emph {et~al.}(2015)\citenamefont
		{Hizanidis}, \citenamefont {Panagakou}, \citenamefont {Omelchenko},
		\citenamefont {Sch{\"o}ll}, \citenamefont {H{\"o}vel},\ and\ \citenamefont
		{Provata}}]{hizanidis2015chimera}%
	\BibitemOpen
	\bibfield  {author} {\bibinfo {author} {\bibfnamefont {J.}~\bibnamefont
			{Hizanidis}}, \bibinfo {author} {\bibfnamefont {E.}~\bibnamefont
			{Panagakou}}, \bibinfo {author} {\bibfnamefont {I.}~\bibnamefont
			{Omelchenko}}, \bibinfo {author} {\bibfnamefont {E.}~\bibnamefont
			{Sch{\"o}ll}}, \bibinfo {author} {\bibfnamefont {P.}~\bibnamefont
			{H{\"o}vel}},\ and\ \bibinfo {author} {\bibfnamefont {A.}~\bibnamefont
			{Provata}},\ }\bibfield  {title} {\bibinfo {title} {Chimera states in
			population dynamics: networks with fragmented and hierarchical
			connectivities},\ }\href@noop {} {\bibfield  {journal} {\bibinfo  {journal}
			{Physical Review E}\ }\textbf {\bibinfo {volume} {92}},\ \bibinfo {pages}
		{012915} (\bibinfo {year} {2015})}\BibitemShut {NoStop}%
	\bibitem [{\citenamefont {Bera}\ \emph {et~al.}(2017)\citenamefont {Bera},
		\citenamefont {Majhi}, \citenamefont {Ghosh},\ and\ \citenamefont
		{Perc}}]{bera2017chimera}%
	\BibitemOpen
	\bibfield  {author} {\bibinfo {author} {\bibfnamefont {B.~K.}\ \bibnamefont
			{Bera}}, \bibinfo {author} {\bibfnamefont {S.}~\bibnamefont {Majhi}},
		\bibinfo {author} {\bibfnamefont {D.}~\bibnamefont {Ghosh}},\ and\ \bibinfo
		{author} {\bibfnamefont {M.}~\bibnamefont {Perc}},\ }\bibfield  {title}
	{\bibinfo {title} {Chimera states: effects of different coupling
			topologies},\ }\href@noop {} {\bibfield  {journal} {\bibinfo  {journal} {EPL
				(Europhysics Letters)}\ }\textbf {\bibinfo {volume} {118}},\ \bibinfo {pages}
		{10001} (\bibinfo {year} {2017})}\BibitemShut {NoStop}%
	\bibitem [{\citenamefont {Meena}\ \emph {et~al.}(2016)\citenamefont {Meena},
		\citenamefont {Murali},\ and\ \citenamefont {Sinha}}]{meena2016chimera}%
	\BibitemOpen
	\bibfield  {author} {\bibinfo {author} {\bibfnamefont {C.}~\bibnamefont
			{Meena}}, \bibinfo {author} {\bibfnamefont {K.}~\bibnamefont {Murali}},\ and\
		\bibinfo {author} {\bibfnamefont {S.}~\bibnamefont {Sinha}},\ }\bibfield
	{title} {\bibinfo {title} {Chimera states in star networks},\ }\href@noop {}
	{\bibfield  {journal} {\bibinfo  {journal} {International Journal of
				Bifurcation and Chaos}\ }\textbf {\bibinfo {volume} {26}},\ \bibinfo {pages}
		{1630023} (\bibinfo {year} {2016})}\BibitemShut {NoStop}%
	\bibitem [{\citenamefont {Sethia}\ and\ \citenamefont
		{Sen}(2014)}]{sethia2014chimera}%
	\BibitemOpen
	\bibfield  {author} {\bibinfo {author} {\bibfnamefont {G.~C.}\ \bibnamefont
			{Sethia}}\ and\ \bibinfo {author} {\bibfnamefont {A.}~\bibnamefont {Sen}},\
	}\bibfield  {title} {\bibinfo {title} {Chimera states: The existence criteria
			revisited},\ }\href@noop {} {\bibfield  {journal} {\bibinfo  {journal}
			{Physical review letters}\ }\textbf {\bibinfo {volume} {112}},\ \bibinfo
		{pages} {144101} (\bibinfo {year} {2014})}\BibitemShut {NoStop}%
	\bibitem [{\citenamefont {Yeldesbay}\ \emph {et~al.}(2014)\citenamefont
		{Yeldesbay}, \citenamefont {Pikovsky},\ and\ \citenamefont
		{Rosenblum}}]{yeldesbay2014chimeralike}%
	\BibitemOpen
	\bibfield  {author} {\bibinfo {author} {\bibfnamefont {A.}~\bibnamefont
			{Yeldesbay}}, \bibinfo {author} {\bibfnamefont {A.}~\bibnamefont
			{Pikovsky}},\ and\ \bibinfo {author} {\bibfnamefont {M.}~\bibnamefont
			{Rosenblum}},\ }\bibfield  {title} {\bibinfo {title} {Chimeralike states in
			an ensemble of globally coupled oscillators},\ }\href@noop {} {\bibfield
		{journal} {\bibinfo  {journal} {Physical review letters}\ }\textbf {\bibinfo
			{volume} {112}},\ \bibinfo {pages} {144103} (\bibinfo {year}
		{2014})}\BibitemShut {NoStop}%
	\bibitem [{\citenamefont {Hens}\ \emph
		{et~al.}(2015{\natexlab{a}})\citenamefont {Hens}, \citenamefont {Mishra},
		\citenamefont {Roy}, \citenamefont {Sen},\ and\ \citenamefont
		{Dana}}]{hens2015chimera}%
	\BibitemOpen
	\bibfield  {author} {\bibinfo {author} {\bibfnamefont {C.}~\bibnamefont
			{Hens}}, \bibinfo {author} {\bibfnamefont {A.}~\bibnamefont {Mishra}},
		\bibinfo {author} {\bibfnamefont {P.}~\bibnamefont {Roy}}, \bibinfo {author}
		{\bibfnamefont {A.}~\bibnamefont {Sen}},\ and\ \bibinfo {author}
		{\bibfnamefont {S.}~\bibnamefont {Dana}},\ }\bibfield  {title} {\bibinfo
		{title} {Chimera states in a population of identical oscillators under planar
			cross-coupling},\ }\href@noop {} {\bibfield  {journal} {\bibinfo  {journal}
			{Pramana}\ }\textbf {\bibinfo {volume} {84}},\ \bibinfo {pages} {229}
		(\bibinfo {year} {2015}{\natexlab{a}})}\BibitemShut {NoStop}%
	\bibitem [{\citenamefont {Mishra}\ \emph {et~al.}(2015)\citenamefont {Mishra},
		\citenamefont {Hens}, \citenamefont {Bose}, \citenamefont {Roy},\ and\
		\citenamefont {Dana}}]{mishra2015chimeralike}%
	\BibitemOpen
	\bibfield  {author} {\bibinfo {author} {\bibfnamefont {A.}~\bibnamefont
			{Mishra}}, \bibinfo {author} {\bibfnamefont {C.}~\bibnamefont {Hens}},
		\bibinfo {author} {\bibfnamefont {M.}~\bibnamefont {Bose}}, \bibinfo {author}
		{\bibfnamefont {P.~K.}\ \bibnamefont {Roy}},\ and\ \bibinfo {author}
		{\bibfnamefont {S.~K.}\ \bibnamefont {Dana}},\ }\bibfield  {title} {\bibinfo
		{title} {Chimeralike states in a network of oscillators under attractive and
			repulsive global coupling},\ }\href@noop {} {\bibfield  {journal} {\bibinfo
			{journal} {Physical Review E}\ }\textbf {\bibinfo {volume} {92}},\ \bibinfo
		{pages} {062920} (\bibinfo {year} {2015})}\BibitemShut {NoStop}%
	\bibitem [{\citenamefont {Zakharova}\ \emph {et~al.}(2016)\citenamefont
		{Zakharova}, \citenamefont {Kapeller},\ and\ \citenamefont
		{Sch{\"o}ll}}]{zakharova2016amplitude}%
	\BibitemOpen
	\bibfield  {author} {\bibinfo {author} {\bibfnamefont {A.}~\bibnamefont
			{Zakharova}}, \bibinfo {author} {\bibfnamefont {M.}~\bibnamefont
			{Kapeller}},\ and\ \bibinfo {author} {\bibfnamefont {E.}~\bibnamefont
			{Sch{\"o}ll}},\ }\bibfield  {title} {\bibinfo {title} {Amplitude chimeras and
			chimera death in dynamical networks},\ }in\ \href@noop {} {\emph {\bibinfo
			{booktitle} {Journal of Physics: Conference Series}}},\ Vol.\ \bibinfo
	{volume} {727}\ (\bibinfo {organization} {IOP Publishing},\ \bibinfo {year}
	{2016})\ p.\ \bibinfo {pages} {012018}\BibitemShut {NoStop}%
	\bibitem [{\citenamefont {Banerjee}\ \emph {et~al.}(2018)\citenamefont
		{Banerjee}, \citenamefont {Biswas}, \citenamefont {Ghosh}, \citenamefont
		{Sch{\"o}ll},\ and\ \citenamefont {Zakharova}}]{banerjee2018networks}%
	\BibitemOpen
	\bibfield  {author} {\bibinfo {author} {\bibfnamefont {T.}~\bibnamefont
			{Banerjee}}, \bibinfo {author} {\bibfnamefont {D.}~\bibnamefont {Biswas}},
		\bibinfo {author} {\bibfnamefont {D.}~\bibnamefont {Ghosh}}, \bibinfo
		{author} {\bibfnamefont {E.}~\bibnamefont {Sch{\"o}ll}},\ and\ \bibinfo
		{author} {\bibfnamefont {A.}~\bibnamefont {Zakharova}},\ }\bibfield  {title}
	{\bibinfo {title} {Networks of coupled oscillators: from phase to amplitude
			chimeras},\ }\href@noop {} {\bibfield  {journal} {\bibinfo  {journal} {Chaos:
				An Interdisciplinary Journal of Nonlinear Science}\ }\textbf {\bibinfo
			{volume} {28}},\ \bibinfo {pages} {113124} (\bibinfo {year}
		{2018})}\BibitemShut {NoStop}%
	\bibitem [{\citenamefont {Zakharova}\ \emph {et~al.}(2014)\citenamefont
		{Zakharova}, \citenamefont {Kapeller},\ and\ \citenamefont
		{Sch{\"o}ll}}]{zakharova2014chimera}%
	\BibitemOpen
	\bibfield  {author} {\bibinfo {author} {\bibfnamefont {A.}~\bibnamefont
			{Zakharova}}, \bibinfo {author} {\bibfnamefont {M.}~\bibnamefont
			{Kapeller}},\ and\ \bibinfo {author} {\bibfnamefont {E.}~\bibnamefont
			{Sch{\"o}ll}},\ }\bibfield  {title} {\bibinfo {title} {Chimera death:
			Symmetry breaking in dynamical networks},\ }\href@noop {} {\bibfield
		{journal} {\bibinfo  {journal} {Physical review letters}\ }\textbf {\bibinfo
			{volume} {112}},\ \bibinfo {pages} {154101} (\bibinfo {year}
		{2014})}\BibitemShut {NoStop}%
	\bibitem [{\citenamefont {Banerjee}(2015)}]{banerjee2015mean}%
	\BibitemOpen
	\bibfield  {author} {\bibinfo {author} {\bibfnamefont {T.}~\bibnamefont
			{Banerjee}},\ }\bibfield  {title} {\bibinfo {title}
		{Mean-field-diffusion--induced chimera death state},\ }\href@noop {}
	{\bibfield  {journal} {\bibinfo  {journal} {EPL (Europhysics Letters)}\
		}\textbf {\bibinfo {volume} {110}},\ \bibinfo {pages} {60003} (\bibinfo
		{year} {2015})}\BibitemShut {NoStop}%
	\bibitem [{\citenamefont {Maistrenko}\ \emph {et~al.}(2017)\citenamefont
		{Maistrenko}, \citenamefont {Brezetsky}, \citenamefont {Jaros}, \citenamefont
		{Levchenko},\ and\ \citenamefont {Kapitaniak}}]{maistrenko2017smallest}%
	\BibitemOpen
	\bibfield  {author} {\bibinfo {author} {\bibfnamefont {Y.}~\bibnamefont
			{Maistrenko}}, \bibinfo {author} {\bibfnamefont {S.}~\bibnamefont
			{Brezetsky}}, \bibinfo {author} {\bibfnamefont {P.}~\bibnamefont {Jaros}},
		\bibinfo {author} {\bibfnamefont {R.}~\bibnamefont {Levchenko}},\ and\
		\bibinfo {author} {\bibfnamefont {T.}~\bibnamefont {Kapitaniak}},\ }\bibfield
	{title} {\bibinfo {title} {Smallest chimera states},\ }\href@noop {}
	{\bibfield  {journal} {\bibinfo  {journal} {Physical Review E}\ }\textbf
		{\bibinfo {volume} {95}},\ \bibinfo {pages} {010203} (\bibinfo {year}
		{2017})}\BibitemShut {NoStop}%
	\bibitem [{\citenamefont {Hart}\ \emph {et~al.}(2016)\citenamefont {Hart},
		\citenamefont {Bansal}, \citenamefont {Murphy},\ and\ \citenamefont
		{Roy}}]{hart2016experimental}%
	\BibitemOpen
	\bibfield  {author} {\bibinfo {author} {\bibfnamefont {J.~D.}\ \bibnamefont
			{Hart}}, \bibinfo {author} {\bibfnamefont {K.}~\bibnamefont {Bansal}},
		\bibinfo {author} {\bibfnamefont {T.~E.}\ \bibnamefont {Murphy}},\ and\
		\bibinfo {author} {\bibfnamefont {R.}~\bibnamefont {Roy}},\ }\bibfield
	{title} {\bibinfo {title} {Experimental observation of chimera and cluster
			states in a minimal globally coupled network},\ }\href@noop {} {\bibfield
		{journal} {\bibinfo  {journal} {Chaos: An Interdisciplinary Journal of
				Nonlinear Science}\ }\textbf {\bibinfo {volume} {26}},\ \bibinfo {pages}
		{094801} (\bibinfo {year} {2016})}\BibitemShut {NoStop}%
	\bibitem [{\citenamefont {Senthilkumar}\ and\ \citenamefont
		{Chandrasekar}(2019)}]{senthilkumar2019local}%
	\BibitemOpen
	\bibfield  {author} {\bibinfo {author} {\bibfnamefont {D.}~\bibnamefont
			{Senthilkumar}}\ and\ \bibinfo {author} {\bibfnamefont {V.}~\bibnamefont
			{Chandrasekar}},\ }\bibfield  {title} {\bibinfo {title} {Local and global
			chimera states in a four-oscillator system},\ }\href@noop {} {\bibfield
		{journal} {\bibinfo  {journal} {Physical Review E}\ }\textbf {\bibinfo
			{volume} {100}},\ \bibinfo {pages} {032211} (\bibinfo {year}
		{2019})}\BibitemShut {NoStop}%
	\bibitem [{\citenamefont {Wojewoda}\ \emph {et~al.}(2016)\citenamefont
		{Wojewoda}, \citenamefont {Czolczynski}, \citenamefont {Maistrenko},\ and\
		\citenamefont {Kapitaniak}}]{wojewoda2016smallest}%
	\BibitemOpen
	\bibfield  {author} {\bibinfo {author} {\bibfnamefont {J.}~\bibnamefont
			{Wojewoda}}, \bibinfo {author} {\bibfnamefont {K.}~\bibnamefont
			{Czolczynski}}, \bibinfo {author} {\bibfnamefont {Y.}~\bibnamefont
			{Maistrenko}},\ and\ \bibinfo {author} {\bibfnamefont {T.}~\bibnamefont
			{Kapitaniak}},\ }\bibfield  {title} {\bibinfo {title} {The smallest chimera
			state for coupled pendula},\ }\href@noop {} {\bibfield  {journal} {\bibinfo
			{journal} {Scientific reports}\ }\textbf {\bibinfo {volume} {6}},\ \bibinfo
		{pages} {1} (\bibinfo {year} {2016})}\BibitemShut {NoStop}%
	\bibitem [{\citenamefont {Mishra}\ \emph {et~al.}(2017)\citenamefont {Mishra},
		\citenamefont {Saha}, \citenamefont {Hens}, \citenamefont {Roy},
		\citenamefont {Bose}, \citenamefont {Louodop}, \citenamefont {Cerdeira},\
		and\ \citenamefont {Dana}}]{mishra2017coherent}%
	\BibitemOpen
	\bibfield  {author} {\bibinfo {author} {\bibfnamefont {A.}~\bibnamefont
			{Mishra}}, \bibinfo {author} {\bibfnamefont {S.}~\bibnamefont {Saha}},
		\bibinfo {author} {\bibfnamefont {C.}~\bibnamefont {Hens}}, \bibinfo {author}
		{\bibfnamefont {P.~K.}\ \bibnamefont {Roy}}, \bibinfo {author} {\bibfnamefont
			{M.}~\bibnamefont {Bose}}, \bibinfo {author} {\bibfnamefont {P.}~\bibnamefont
			{Louodop}}, \bibinfo {author} {\bibfnamefont {H.~A.}\ \bibnamefont
			{Cerdeira}},\ and\ \bibinfo {author} {\bibfnamefont {S.~K.}\ \bibnamefont
			{Dana}},\ }\bibfield  {title} {\bibinfo {title} {Coherent libration to
			coherent rotational dynamics via chimeralike states and clustering in a
			josephson junction array},\ }\href@noop {} {\bibfield  {journal} {\bibinfo
			{journal} {Physical Review E}\ }\textbf {\bibinfo {volume} {95}},\ \bibinfo
		{pages} {010201} (\bibinfo {year} {2017})}\BibitemShut {NoStop}%
	\bibitem [{\citenamefont {Ray}\ \emph {et~al.}(2020)\citenamefont {Ray},
		\citenamefont {Mishra}, \citenamefont {Ghosh}, \citenamefont {Kapitaniak},
		\citenamefont {Dana},\ and\ \citenamefont {Hens}}]{ray2020extreme}%
	\BibitemOpen
	\bibfield  {author} {\bibinfo {author} {\bibfnamefont {A.}~\bibnamefont
			{Ray}}, \bibinfo {author} {\bibfnamefont {A.}~\bibnamefont {Mishra}},
		\bibinfo {author} {\bibfnamefont {D.}~\bibnamefont {Ghosh}}, \bibinfo
		{author} {\bibfnamefont {T.}~\bibnamefont {Kapitaniak}}, \bibinfo {author}
		{\bibfnamefont {S.~K.}\ \bibnamefont {Dana}},\ and\ \bibinfo {author}
		{\bibfnamefont {C.}~\bibnamefont {Hens}},\ }\bibfield  {title} {\bibinfo
		{title} {Extreme events in a network of heterogeneous josephson junctions},\
	}\href@noop {} {\bibfield  {journal} {\bibinfo  {journal} {Physical Review
				E}\ }\textbf {\bibinfo {volume} {101}},\ \bibinfo {pages} {032209} (\bibinfo
		{year} {2020})}\BibitemShut {NoStop}%
	\bibitem [{\citenamefont {Dana}\ \emph {et~al.}(2006)\citenamefont {Dana},
		\citenamefont {Sengupta},\ and\ \citenamefont {Hu}}]{dana2006spiking}%
	\BibitemOpen
	\bibfield  {author} {\bibinfo {author} {\bibfnamefont {S.~K.}\ \bibnamefont
			{Dana}}, \bibinfo {author} {\bibfnamefont {D.~C.}\ \bibnamefont {Sengupta}},\
		and\ \bibinfo {author} {\bibfnamefont {C.-K.}\ \bibnamefont {Hu}},\
	}\bibfield  {title} {\bibinfo {title} {Spiking and bursting in josephson
			junction},\ }\href@noop {} {\bibfield  {journal} {\bibinfo  {journal} {IEEE
				Transactions on Circuits and Systems II: Express Briefs}\ }\textbf {\bibinfo
			{volume} {53}},\ \bibinfo {pages} {1031} (\bibinfo {year}
		{2006})}\BibitemShut {NoStop}%
	\bibitem [{\citenamefont {Mishra}\ \emph {et~al.}(2021)\citenamefont {Mishra},
		\citenamefont {Ghosh}, \citenamefont {Kumar~Dana}, \citenamefont
		{Kapitaniak},\ and\ \citenamefont {Hens}}]{mishra2021neuron}%
	\BibitemOpen
	\bibfield  {author} {\bibinfo {author} {\bibfnamefont {A.}~\bibnamefont
			{Mishra}}, \bibinfo {author} {\bibfnamefont {S.}~\bibnamefont {Ghosh}},
		\bibinfo {author} {\bibfnamefont {S.}~\bibnamefont {Kumar~Dana}}, \bibinfo
		{author} {\bibfnamefont {T.}~\bibnamefont {Kapitaniak}},\ and\ \bibinfo
		{author} {\bibfnamefont {C.}~\bibnamefont {Hens}},\ }\bibfield  {title}
	{\bibinfo {title} {Neuron-like spiking and bursting in josephson junctions: A
			review},\ }\href@noop {} {\bibfield  {journal} {\bibinfo  {journal} {Chaos:
				An Interdisciplinary Journal of Nonlinear Science}\ }\textbf {\bibinfo
			{volume} {31}},\ \bibinfo {pages} {052101} (\bibinfo {year}
		{2021})}\BibitemShut {NoStop}%
	\bibitem [{\citenamefont {Hongray}\ \emph {et~al.}(2015)\citenamefont
		{Hongray}, \citenamefont {Balakrishnan},\ and\ \citenamefont
		{Dana}}]{hongray2015bursting}%
	\BibitemOpen
	\bibfield  {author} {\bibinfo {author} {\bibfnamefont {T.}~\bibnamefont
			{Hongray}}, \bibinfo {author} {\bibfnamefont {J.}~\bibnamefont
			{Balakrishnan}},\ and\ \bibinfo {author} {\bibfnamefont {S.~K.}\ \bibnamefont
			{Dana}},\ }\bibfield  {title} {\bibinfo {title} {Bursting behaviour in
			coupled josephson junctions},\ }\href@noop {} {\bibfield  {journal} {\bibinfo
			{journal} {Chaos: An Interdisciplinary Journal of Nonlinear Science}\
		}\textbf {\bibinfo {volume} {25}},\ \bibinfo {pages} {123104} (\bibinfo
		{year} {2015})}\BibitemShut {NoStop}%
	\bibitem [{\citenamefont {Hens}\ \emph
		{et~al.}(2015{\natexlab{b}})\citenamefont {Hens}, \citenamefont {Pal},\ and\
		\citenamefont {Dana}}]{hens2015bursting}%
	\BibitemOpen
	\bibfield  {author} {\bibinfo {author} {\bibfnamefont {C.}~\bibnamefont
			{Hens}}, \bibinfo {author} {\bibfnamefont {P.}~\bibnamefont {Pal}},\ and\
		\bibinfo {author} {\bibfnamefont {S.~K.}\ \bibnamefont {Dana}},\ }\bibfield
	{title} {\bibinfo {title} {Bursting dynamics in a population of oscillatory
			and excitable josephson junctions},\ }\href@noop {} {\bibfield  {journal}
		{\bibinfo  {journal} {Physical Review E}\ }\textbf {\bibinfo {volume} {92}},\
		\bibinfo {pages} {022915} (\bibinfo {year} {2015}{\natexlab{b}})}\BibitemShut
	{NoStop}%
	\bibitem [{\citenamefont {Dana}\ \emph {et~al.}(2001)\citenamefont {Dana},
		\citenamefont {Sengupta},\ and\ \citenamefont {Edoh}}]{dana2001chaotic}%
	\BibitemOpen
	\bibfield  {author} {\bibinfo {author} {\bibfnamefont {S.~K.}\ \bibnamefont
			{Dana}}, \bibinfo {author} {\bibfnamefont {D.~C.}\ \bibnamefont {Sengupta}},\
		and\ \bibinfo {author} {\bibfnamefont {K.~D.}\ \bibnamefont {Edoh}},\
	}\bibfield  {title} {\bibinfo {title} {Chaotic dynamics in josephson
			junction},\ }\href@noop {} {\bibfield  {journal} {\bibinfo  {journal} {IEEE
				Transactions on Circuits and Systems I: Fundamental Theory and Applications}\
		}\textbf {\bibinfo {volume} {48}},\ \bibinfo {pages} {990} (\bibinfo {year}
		{2001})}\BibitemShut {NoStop}%
	\bibitem [{\citenamefont {Josephson}(1962)}]{josephson1962possible}%
	\BibitemOpen
	\bibfield  {author} {\bibinfo {author} {\bibfnamefont {B.}~\bibnamefont
			{Josephson}},\ }\bibfield  {title} {\bibinfo {title} {Possible new effect in
			superconducting tunneling},\ }\href@noop {} {\bibfield  {journal} {\bibinfo
			{journal} {Phys. Lett.}\ }\textbf {\bibinfo {volume} {1}},\ \bibinfo {pages}
		{251} (\bibinfo {year} {1962})}\BibitemShut {NoStop}%
	\bibitem [{\citenamefont {Pecora}\ and\ \citenamefont
		{Carroll}(1998)}]{pecora1998master}%
	\BibitemOpen
	\bibfield  {author} {\bibinfo {author} {\bibfnamefont {L.~M.}\ \bibnamefont
			{Pecora}}\ and\ \bibinfo {author} {\bibfnamefont {T.~L.}\ \bibnamefont
			{Carroll}},\ }\bibfield  {title} {\bibinfo {title} {Master stability
			functions for synchronized coupled systems},\ }\href@noop {} {\bibfield
		{journal} {\bibinfo  {journal} {Physical review letters}\ }\textbf {\bibinfo
			{volume} {80}},\ \bibinfo {pages} {2109} (\bibinfo {year}
		{1998})}\BibitemShut {NoStop}%
\end{thebibliography}

%

\end{document}